\documentclass[amssymb,preprint,tightenlines,aps]{revtex4}
\newtheorem{theorem}{Theorem}
\newtheorem{acknowledgement}[theorem]{Acknowledgement}

\begin{document}
\title{Topology, connectivity and electronic structure of C and B cages and the
corresponding nanotubes}
\author{F.E. Leys$^{a}$, C. Amovilli$^{b}$ and N.H. March$^{a,c}$}
\address{$^{a}$Department of Physics, University of Antwerp (RUCA), Groenenborgerlaan%
\\
171 B-2020 Antwerpen, Belgium.\\
$^{b}$Dipartimento di Chimica e Chimica Industriale, Universit\`{a} di Pisa,%
\\
Via Risorgimento 35, 56126 Pisa, Italy\\
$^{c}$Oxford University, Oxford, England.}
\date{\today}


\begin{abstract}
After a brief discussion of the structural trends which appear with
increasing number of atoms in B cages, a one-to one correspondence between
the connectivity of $B$ cages and $C$ cage structures will be proposed. The
electronic level spectra of both systems from Hartree-Fock calculations is
given and discussed. The relation of curvature introduced into an originally
planar graphitic fragment to pentagonal 'defects' such as are present in
buckminsterfullerene is also briefly treated.

A study of the structure and electronic properties of $B$ nanotubes will
then be introduced. We start by presenting a solution of the free-electron
network approach for a 'model boron' planar lattice with local coordination
number $6$. In particular the dispersion relation $E({\bf k})$ for the $\pi
- $electron bands, together with the corresponding electronic Density Of
States ($DOS$), will be exhibited. This is then used within the zone folding
scheme to obtain information about the electronic $DOS$ of different
nanotubes obtained by folding this model boron sheet.

To obtain the self-consistent potential in which the valence electrons move
in a nanotube, 'the March model' in its original form was invoked and
results are reported for a carbon nanotube.

Finally, heterostructures, such as $BN$ cages and fluorinated
buckminsterfullerene, will be briefly treated, the new feature here being
electronegativity difference.
\end{abstract}

\maketitle

\section{Background and outline}

The continuing usefulness of models of $\pi -$electrons in conjugated
systems, for example that of H\"{u}ckel \cite{Huck}, testifies to the
importance of geometry and connectivity in determining electronic structure.
There has, indeed, been renewed interest in this area, due to the potential
for technology of nanostructures \cite{nano}.

Therefore, in the present study of $C$ and $B$ cages, and the corresponding
nanotubes, we shall not hesitate in presenting the simplest possible
approaches to electronic structure of the $\pi -$electrons when these
highlight the importance of topology and connectivity. However, it will also
prove useful in the course of the discussion, to refer briefly to
calculations of Hartree-Fock quality that have been carried out on a variety
of B \cite{boroncl} and C \cite{carboncl} cages.

The outline of the present study is then as follows. Section II below
exposes, essentially via Euler's theorem, a one-to-one correspondence
between the connectivity of $C$ and $B$ cages. For example, $C_{60}\;$%
naturally leads to a $B_{32}$ cage. Considerable similarities between the
electronic level spectra of both systems are reported. Also briefly
discussed is the matter of pentagonal 'defects' and the approximate relation
to curvature of originally planar graphitic fragments. This is illustrated
by reference to the recent Hartree-Fock calculations on $B$ cages of
intermediate size already mentioned.

This discussion involving planar fragments, leads into Section III which
presents a study of the structure and the electronic properties of boron
nanotubes obtained by folding planar sheets of boron atoms. We utilize a
(quantum) model akin to Kirchoff's laws of electrical circuits, where
evidently connectivity is again an essential ingredient, $\pi -$electrons
being constrained to move along bonds (wires) joining neighbouring nuclei.
The dispersion relation of the $\pi -$bands is then utilized to calculate ($%
a $) constant energy surfaces and ($b$) the electronic $DOS$ for this
two-dimensional ($2D$) model boron lattice. This latter quantity is compared
and contrasted with the study of graphene made in the early work of\ Coulson
\cite{Coulson}. These dispersion relations are then used to obtain the
one-dimensional ($1D$) energy bands and the electronic $DOS$ of boron
nanotubes within the 'zone folding scheme' \cite{nano}.

In section IV the theory of the inhomogeneous electron gas \cite{Lund}
is applied self-consistently to infinite nanotubes, but now in a surface
charge model generalizing the 'spherical March model' of C$_{60}$ \cite{Clou,desp,Sir},
originally designed to deal with tetrahedral (eg.
$SiH_{4}\;$and$\;GeH_{4})$ and octahedral molecules \cite{orima}. It is
noteworthy in the present context that central to the March model is the
(obviously approximate) assumption that the $\pi -$electrons (one per $C$
atom) in $C_{60}\;$are distributed spherically. Naturally this is consistent
with the uniform surface charge model. However, with this assumption, if we
assume the nuclei are constrained on the surface of a sphere, the lowest
energy isomer structure will be determined by minimizing the Coulomb
repulsion energy between the nuclei. This is a very old problem going back
as far as J.J. Thomson. The history is briefly recorded by Berezin \cite{probl}
together with other possible references. For general $N$ point
charges the problem remains mathematically unsolved except for small $N$. In
particular exact solutions are known for $N=4,\;6$ and $12$, and are
respectively tetrahedron, octahedron and icosahedron. For $C_{60}\;$a Monte
Carlo programme written specifically for boron cages \cite{boroncl}
demonstrated to numerical accuracy that the European football was extremely
close to (if not lowest) the favoured isomer for this molecule, provided
nuclei are constrained on the surface of the sphere.

The above discussions are then generalized in Section V to treat
hetero-nanostructures, earlier work on $BN$ cages \cite{Zhu} being
briefly summarized, followed by reference to fluorinated buckminsterfullerene
\cite{fluor}. The essential new feature in both examples is the
electronegativity difference between the component atoms. Section VI
constitutes a summary, with some proposals for future directions; the
salient one being the use of the progress on quantum current network models
(see especially the study of Ringwood \cite{Ring}) to investigate
further topologically disordered networks, which began with the work of
Dancz et al. \cite{Dancz}.

\section{Curvature of graphitic fragment, pentagonal defects and a
one-to-one correspondence between connectivity of B and C cages}

\subsection{Use of Euler's theorem}

Amovilli and March \cite{boroncl} performed Hartree-Fock calculations on
$B_{n}\;$cages, with $n$ varying from $30$ to $54$,$\;$under the constraint
that all $B$ nuclei lie on the surface of a sphere. In correspondence with
the 'Aufbau principle' all boron atoms were found to be either $5$- or $6$%
-fold coordinated. They noted that as the number of boron atoms gets large,
the number of triangular faces becomes dominant compared to other types of
faces in the optimized geometry and the number of penta-coordinated atoms
approaches $\sim 12$ for the largest cages, though considerable scatter
around this result was found.

In fact, if one assumes {\it all} faces to be triangular, one can rigorously
show that the number of penta-coordinated atoms is {\it exactly} $12$, as we
demonstrate immediately below. If all boron faces are triangular we have the
following relation between the number of boron faces $f_{B}\;$and the number
of boron edges $e_{B}$%
\begin{equation}
e_{B}=\frac{3}{2}f_{B}  \label{tricon}
\end{equation}
Following the aufbau principle we only have $5-$ and $6-$fold coordinated
boron atoms and so if$\;n_{5}\;$and$\;n_{6}$ denote the number of $5$- and $%
6 $-fold coordinated boron atoms respectively we have that
\begin{equation}
e_{B}=\frac{5}{2}n_{5}+\frac{6}{2}n_{6}  \label{56}
\end{equation}
with the total number of boron atoms $n$ evidently given by
\begin{equation}
n=n_{5}+n_{6}  \label{nn5n6}
\end{equation}
But now for any closed geometrical figure, the number of faces $f$, the
number of edges $e$, and the number of vertices $v$ are related through
Euler's theorem
\[
f+v=e+2
\]
and\ combining\ this\ theorem, where obvioulsy the number of boron vertices $%
v_{B}=n$,\ with Eqs.(\ref{tricon}),\ (\ref{56})\ and (\ref{nn5n6})\ we\
immediately obtain\ that\ $n_{5}=12$\ i.e. the\ number of $5$-fold
coordinated atoms in boron clusters is uniquely predicted to be $12$ if one
assumes all faces to be triangular.

This is reminiscent of the situation for carbon fullerenes where, as was
already noted by Euler, to form a closed figure with hexagons, one always
needs exactly $12$ pentagons. This similarity inspired us to examine whether
a correspondence (or map) between the structure of a $B_{n}\;$boron cage,
containing only triangular faces, and the structure of a $C_{N}$ fullerene,
containing only pentagons and hexagons, can be established.

As will be demonstrated below, the only constraint one needs to impose is
that every $B-B$ bond crosses exactly one $C-C$ bond. Evidently this implies
that there is only one carbon atom in every boron face and vice versa.

First of all, since every carbon atom in a $C_{N}$ cage is threefold
coordinated through $sp^{2}\;$hybridization, the crossing of the bonds is a
sufficient condition to ensure that all boron faces are triangular as is
required. Secondly, as is clear from Fig. $2.1$, a pentagonal 'defect' in a $%
C_{N}$ cage (a pentagon surrounded by $5$ hexagons, see Fig. $2.1$, the thin
lines) will generate a corresponding pentagonal 'defect' in a boron cage\ (a
penta-coordinated atom surrounded by $5$ hexa-coordinated atoms, see Fig. $%
2.1$, the thick lines). On the other hand, as one can easily check, sections
of the $C_{N}$ cage containing only hexagons will generate only $6$-fold
coordinated boron atoms.

So a natural correspondence between the structure and connectivity of $C_{N}$
\ carbon cages and that of $B_{n}\;$boron cages (with only triangular faces)
is exposed. To find out precisely which $B_{n}\;$boron cage we obtain in
this way starting from a certain $C_{N}\;$cage we again turn to Euler's
theorem. Starting from a carbon cage $C_{N},\;$we have obviously $N$
vertices, and hence
\begin{equation}
f_{c}=e_{c}+2-N  \label{eulc}
\end{equation}

With constant threefold coordination, it follows that the number of edges, $%
e_{c},$ is equal to $\frac{3}{2}N\;$and hence, inserting this value into Eq.(%
\ref{eulc}) one reaches the result for $C_{N}\;$cages that
\begin{equation}
f_{c}=\frac{N}{2}+2  \label{eulcr}
\end{equation}
Since our construction implies one boron atom in every carbon face, the
number of carbon faces $f_{c}\;$obviously equals the number of boron
vertices $v_{B}$ (=$n$)$\;$and we obtain the following one-to-one relation
between $N$ and $n$.
\begin{equation}
n=\frac{N}{2}+2  \label{conrel}
\end{equation}
Eq.(\ref{conrel}) applied to buckminsterfullerene itself with $N=60$ yields $%
n=32$, i.e. the cage $B_{32}\;$is its analogue. Note that this
correspondence is purely topological. Whether or not realistic values for
the bond lengths in $C_{N}$ \ carbon cages and $B_{n}\;$boron cages allow
for the mapping we propose in Fig. $2.1$ is not relevant. Since the faces in
$B_{n}\;$boron cages\ become predominantly triangular only in the limit of
large $n$, we expect the correspondence to become\ especially relevant for
large $B_{n}\;$boron clusters.

\subsection{Electronic structure of $\protect\pi -$electrons in $C_{60}\;$%
and valence electrons in $B_{32}$}

Having emphasized topology and connectivity above, let us first record the
one-electron eigenvalues obtained by Amovilli et al. \cite{carboncl}
from Hartree-Fock calculations for $C_{60}$. To emphasize the degeneracies
plus near degeneracies , the reader may refer to Fig. $2.2$ As is readily
verified, filling these depicted $\pi $-eigenvalues with two electrons per
level shows that the upper state is the {\it HOMO} level.

Turning to $B_{32},$ a large number of eigenvalues for the valence orbitals
were obtained in the Hartree-Fock study of Amovilli and March
   \cite{boroncl}.
Their degeneracies (or near degeneracies) are depicted in Fig. $%
2.3$. Filling the lowest four graphs of levels would deal with $B_{32}$
itself if one assumes one $\pi $-electron per atom, which is a chemical
oversimplification however.

\subsection{Pentagonal defects and curvature}

To complete the present section, we wish to add some comments on cage
curvature in relation to pentagonal 'defects' (see also ref.\cite{carboncl}%
). If we discuss fullerenes in particular, we first note that there will be
appreciable anisotropic curvature at the equilibrium geometries. To
illustrate this, a best candidate for uniform curvature is a fullerene with
the pentagonal 'defects' spread as uniformly as is feasible: i.e. to form an
icosahedral-symmetry cage. For such a fullerene (which we conceive for the $%
C_{N}\;$cage in the large $N$ limit) the pentagons can be visualized to
reside at the corners of an icosahedron-type super-structure, all the
remaining rings being hexagons. Then it is helpful to consider a graphitic
region corresponding to a triangular face of the above super-structure, all
the remaining rings being hexagons. The area of this face as measured on the
surface of a uniformly curved spherical surface of radius $R$ is evidently $%
A_{curved}=(4\pi R^{2})/20,$ there being $20$ such faces. The length $l$ of
the side of such a 'spherical triangle' is $\theta R;$ here $\theta $
denotes the angle subtended by the edge as viewed from the center of the
sphere. The area of a planar triangle with edges of the same length is $%
A_{plane}=(l/2)(3/4l)^{\frac{1}{2}}.\;$But $A_{plane}$ (corresponding to an
unstrained part of graphene) is different, of course, from $A_{curved},\;$%
and there is a strain $\sigma \;$per $C$ atom given by \cite{carboncl}:
\begin{equation}
\sigma \sim \left\{ A_{plane}/(N/20)\right\} ^{\frac{1}{2}}-\left\{
A_{curved}/(N/20)\right\} ^{\frac{1}{2}}  \label{stress}
\end{equation}
such a strain occuring for each bond in the graphitic portion of the
network. Therefore, with $\sim 3N/2\;$such bonds, the total stress is
\begin{equation}
E_{stress}\approx (3N/2)\frac{1}{2}k\sigma ^{2}  \label{enerstress}
\end{equation}
where $k$ denotes a suitable force constant appropriate for $C-C$ aromatic
bonds. Substituting Eq.(\ref{stress}) into Eq.(\ref{enerstress}) it follows
that
\begin{equation}
E_{stress}\sim 15k\left( \frac{3^{\frac{1}{4}}}{2}\theta -\left\{ \frac{\pi
}{5}\right\} ^{\frac{1}{2}}\right) R^{2}  \label{stressend}
\end{equation}
But the equilibrium R is determined well by \ the 'rule' of constant surface
area per atom \cite{boroncl,carboncl} i.e. $R^{2}\sim N$ and so
from Eq.(\ref{stressend}) stress is important. If some structure other than
that of icosahedral symmetry is assumed for the arrangement of the
pentagons, the geometric factor in Eq.(\ref{stressend}) will be modified,
but will remain non-zero.

The question as to the relief of such curvature strain then arises (still
one is considering\ the large $N$ limit). Once more, as a prototypical
example, the arrangement of icosahedral symmetry can be taken. One then
visualizes the structure to deform to resemble a icosahedron
super-structure, having the pentagon at its apices. Then the triangular
graphitic regions already discussed can be viewed as changed from their
spherically curved forms to almost planar regions with anisotropic curvature
at the edges which connect each triangular region to adjacent ones. Within
the planar triangular areas there is neither strain nor stress: instead it
occurs at the edges of the triangles. The total amount of such edge is $\sim
30R\propto N^{\frac{1}{2}},\;$so that the stress is reduced to $%
E_{stress}^{\prime }\propto N^{\frac{1}{2}}.$ The proportionality constant
can again be expected to depend on the way the pentagons are dispersed, the
overall conclusion being that geometric Gaussion curvature is preferably
localized in the region of the pentagons, with then some anisotropic
curvature (without Gaussian curvature) mediating between such adjacent
parts, as noted in ref.\cite{carboncl}. Computations on large non-open-shell
icosahedral-symmetry fullerenes reveal the proposed polyhedralization in
diagrams of the geometry optimized structures (most clearly for $C_{240};\;$%
the largest treated).

Having referred in some detail above to the graphitic layers, we follow this
account by taking a planar 'model boron' structure and using again topology
and connectivity to discuss the $\pi -$electronic energy band structure. As
was reported in the paper by Boustani et al. \cite{Bous} preliminary
experimental results for pure-boron systems seem to confirm the existence of
boron sheets. We will consider here a purely planar boron configuration and
apply the simplest possible theories to extract the main features of the $%
DOS\;$which arise from topological aspects.

\section{Electronic properties of boron nanotubes.}

\subsection{Structure of a single wall boron nanotube}

\subsubsection{Equilateral and Isosceles zigzag symmetry.}

We start by describing the model planar boron lattice which will then be
folded to obtain boron nanotubes. The structure of the boron lattice we
consider can be obtained simply by considering an ordinary sheet of graphene
in which we replace all carbon atoms by boron atoms and place an additional
boron atom in the center of every hexagon. In this way, every boron atom has
a near-neighbour coordination, say $c$, equal to $6$. The resulting lattice
is shown in Fig. $3.1$

To describe this lattice one can use two lattice vectors, say ${\bf a}_{1}$
and ${\bf a}_{2}$, which are either $120%
%TCIMACRO{\UNICODE[m]{0xb0}}%
%BeginExpansion
{{}^\circ}%
%EndExpansion
\;$apart, or $60%
%TCIMACRO{\UNICODE[m]{0xb0}}%
%BeginExpansion
{{}^\circ}%
%EndExpansion
\;$apart. If one only wants to describe the $2D$ lattice, both choices are
of course equivalent. However, if one wants to use these lattice vectors to
describe the nanotubes obtained by folding the $2D$ sheet, then the choice
where both lattice vectors are $60%
%TCIMACRO{\UNICODE[m]{0xb0}}%
%BeginExpansion
{{}^\circ}%
%EndExpansion
\;$apart is the most convenient as became apparent for carbon nanotubes.

To describe a nanotube obtained by folding a sheet, one starts by defining
the so-called chiral vector ${\bf C}_{h}.\;$
\begin{equation}
{\bf C}_{h}=n{\bf a}_{1}+m{\bf a}_{2}  \label{ch}
\end{equation}
This vector connects by definition two lattice points on the sheet
which have to be connected when one folds the sheet into a tube
and the specific choice for $n$ and $m$ completely determines the
structure of the resulting tube. From the definition of the chiral
vector ${\bf C}_{h}$ it is clear that its length equals the
circumference $L$ of the resulting nanotube which is then given by
\begin{equation}
L=a\sqrt{n^{2}+m^{2}+nm}  \label{length}
\end{equation}
with$\;a=\left| {\bf a}_{1}\right| =\left| {\bf a}_{2}\right| .$

Similar to carbon, all possible distinct seamless nanotubes can be obtained
by choosing a set of indices ($n,m$) with $n$ any integer number and $m$
ranging from $0$ to $n$. However, an interesting feature then arises when
one is dealing with boron. As one can easily see on Fig. $3.1$\ both the
nanotube generated by a ${\bf C}_{h}(n,n)$ chiral vector and that generated
by the ${\bf C}_{h}(n,0)$ vector now have in essence zigzag symmetry.
However, as is clear from Fig. $3.1$ in the case $n=m\;$we have isosceles
triangles{\it \ (}two equal sides), while in the case $m=0$ we have {\it %
equilateral triangles (}three sides equal) triangles. For
simplicity we will
refer to these limiting cases as $(n,n)$ {\it i-zigzag} and $(n,0)$ {\it %
e-zigzag} symmetry respectively where '$i$'$\;$then evidently stands for
isosceles and '$e$' for equilateral.

We will now go on to define the unit cell of our boron nanotube and the
corresponding Brillouin zone. Except for a few details this is identical to
the carbon case and we refer the reader to the excellent book by Saito,
Dresselhaus and Dresselhaus \cite{nano} for more details.

The vector perpendicular to ${\bf C}_{h}\;$going from the chosen origin to
the nearest lattice point defines the translational vector of our $1D$
periodic nanotube and is given by.
\[
{\bf T=}t_{1}{\bf a}_{1}+t_{2}{\bf a}_{2}
\]
with the relation between $t_{1},t_{2}\;$and $m,n$ unchanged\ as compared to
the carbon system namely
\begin{equation}
t_{1}=\frac{2m+n}{d_{R}},\;t_{2}=-\frac{2n+m}{d_{R}}
\end{equation}
where $d_{R}\;$is the greatest common divisor $(gcd)$ of $(2m+n)$ and $%
(2n+m).\;$The rectangle generated by these two vectors ${\bf C}_{h}\;$and $%
{\bf T}$\ is the unit cell of our boron nanotube, where the translational
vector ${\bf T}$ determines the direction in which the unit cell repeats
itself periodically.

\subsubsection{Reciprocal space}

The reciprocal space of the boron sheet is generated by the reciprocal
lattice vectors ${\bf b}_{1}\;$and$\;{\bf b}_{2}\;$defined by
\[
{\bf b}_{i}.{\bf a}_{i}=2\pi \delta _{ij}
\]
We define the vectors in the reciprocal space of the nanotube by the
relations
\begin{eqnarray}
{\bf C}_{h}.{\bf K}_{1} &=&2\pi \;\;\;\;\;{\bf T.K}_{1}=0  \label{reci1} \\
{\bf C}_{h}.{\bf K}_{2} &=&0\;\;\;\;\;\ \ {\bf T.K}_{2}=2\pi  \label{reci2}
\end{eqnarray}
which then immediately leads to the result
\begin{equation}
{\bf K}_{1}=\frac{1}{N}(-t_{2}{\bf b}_{1}+t_{1}{\bf b}_{2})\;\;\;\;\;\;\;\;%
\;\;\;\ \ \ {\bf K}_{2}=\frac{1}{N}(m{\bf b}_{1}-n{\bf b}_{2})  \label{ks}
\end{equation}
with
\begin{equation}
N=mt_{1}-nt_{2}=\frac{\left| {\bf C}_{h}\times {\bf T}\right| }{\left| {\bf a%
}_{1}\times {\bf a}_{2}\right| }=\frac{2(m^{2}+n^{2}+nm)}{d_{R}}  \label{D}
\end{equation}
with ${\bf b}_{1}\;$and ${\bf b}_{2}\;$the lattice vectors of the graphene
Brillouin zone and $N\;$the number of atoms in the unit cell of the
nanotube. Note that for the carbon sheet the number of carbon atoms in the
nanotube unit cell was given by $2N$.

Since the boron lattice we consider can be obtained from taking only the
{\it lattice} points in a graphene layer (points with identical physical
environment), the Brillouin zone of the boron and the carbon sheet have
identical symmetries.

\subsubsection{Energy bands within the 'zone folding sheme'}

Within the 'zone-folding sheme' \cite{nano} one essentially assumes that
the motion of the electrons in a real single wall $3D$ nanotube can be
described by the motion of the electrons in the planar $2D$ strip from which
that specific nanotube can be obtained by folding. In order to obtain the
energy dispersion relations for the boron nanotube we first need to
determine the energy dispersion relation of the boron sheet, say $E_{2DB}(%
{\bf k),}$ for all possible ${\bf k-}$values and then restrict ourselves
only to those ${\bf k-}$values which are consistent with the boundary
conditions for motion of the electrons on the strip which generates the
boron nanotube.

Since the unit cell of the nanotube (or the infinitely long strip) defined
by ${\bf T\;}$and ${\bf C}_{h}\;$repeats itself periodically only in one
dimension (in the direction of ${\bf T)}$ the motion of the electrons along
that direction is characterized in ${\bf k-}$space\ by a quasi-continuous
scalar wave-vector $k$ in a one-dimensional Brillouin zone of length $\frac{%
2\pi }{T}$ in the direction of ${\bf K}_{2}.$ The boundary conditions along
the direction of the chiral vector ${\bf C}_{h}$ yield a discrete number
(equal to $N$) of allowed values for the component of the wave-vector in the
direction of ${\bf K}_{1}$ given by$\;\mu \left( \frac{2\pi }{\left| {\bf C}%
_{h}\right| }\right) \;\left( \text{or\ equivalently\ }\mu \left| {\bf K}%
_{1}\right| \right) \;$with $\mu =0,.....,N-1$.

Combining this we find that the allowed ${\bf k-}$values lie on a set of $N$
parallel lines in ${\bf k-}$space directed along ${\bf K}_{2}$ and separated
by ${\bf K}_{1}$. Every line represents the $1D$ Brillouin zone but for a
different value of the component of the wave-vector along ${\bf K}_{1}.\;$%
The energy values for these wave-vectors are obtained by taking cross
sections along these lines of the energy surface of the full $2D$ boron
sheet. Every cross-section then gives a band in the $1D$ Brillouin zone of
the nanotube.

We then find for the energy dispersion relations for a single wall nanotube
within the 'zone folding scheme' \cite{nano}
\begin{equation}
E_{\mu }(k)=E_{2DB}\left( k\frac{{\bf K}_{2}}{\left| {\bf K}_{2}\right| }%
+\mu {\bf K}_{1}\right) \;\ \ \left( -\frac{\pi }{T}<k<\frac{\pi }{T}\right)
\end{equation}
with the band index $\mu $ given by
\begin{equation}
\mu =0,.....,N-1  \label{bandindex}
\end{equation}
and with $T=\left| {\bf T}\right| .\;$We now first turn to the
determintation of the energy dispersion relation $E_{2DB}({\bf k)\;}$for the
planar boron sheet.

\subsection{Electronic structure of the 2D boron sheet}

\subsubsection{2D Energy dispersion bands}

For the energy band of the $2D$ boron sheet we obtain within the Quantum
Network ($QN$) (See Appendix A) model the relation
\begin{equation}
E_{2DB}^{QN}({\bf k)}=\frac{q^{2}}{2}=\frac{1}{2}\left\{ \frac{1}{a}\cos
^{-1}\left( \frac{1}{c}S({\bf k)}\right) \right\} ^{2}  \label{QN}
\end{equation}
The structure factor $S({\bf k)\;}$appearing in Eq.(\ref{QN}) is defined by
\begin{equation}
S({\bf k)=}\stackrel{c}{%
%TCIMACRO{\underset{j=1}{\sum }}%
%BeginExpansion
\mathrel{\mathop{\sum }\limits_{j=1}}%
%EndExpansion
}e^{i{\bf k}.{\bf a}_{j}}  \label{struct}
\end{equation}
where the vectors ${\bf a}_{j}$ connect every lattice position ${\bf R}\;$to
its $c$ nearest neighbours. On the other hand, from the tight binding ($TB$)
method we obtain the relation (see Appendix B)
\begin{equation}
E_{2DB}^{TB}({\bf k)=}\epsilon _{\pi }+tS({\bf k)}  \label{tb}
\end{equation}
with $\epsilon _{\pi }\;$and $t$ defined in Appendix B. Since we are mainly
interested in qualitative electronic properties which arise as a consequence
of the periodicity and topology of the system under consideration, we have
assumed in our tight binding calculations the following simple values
(assumed arbitrarily in atomic units) for the parameters.
\begin{equation}
\epsilon _{\pi }=0\;,\;t=-1\;\;\;(a.u.)  \label{simppar}
\end{equation}

The important point to stress here is that both expressions for the energy
depend on the wave-vector ${\bf k}$ through the same structure factor $S(%
{\bf k)}$ for which we will derive an explicit expression immediately below.

\subsubsection{Expression for the structure factor S(k) within a given
reference frame}

In particular we start by choosing the $x-y$ coordinate system in the same
direction relative to the lattice vectors ${\bf a}_{1}\;$and ${\bf a}_{2}\;$%
as in Fig. $3.1$. This then yields for the coordinates of both lattice
vectors
\begin{equation}
{\bf a}_{1}=\left( \frac{\sqrt{3}}{2}a\;,\;\frac{a}{2}\right) ,\;\;\;{\bf a}%
_{2}=\left( \frac{\sqrt{3}}{2}a\;,\;-\frac{a}{2}\right)  \label{lattco}
\end{equation}
where the lattice constant $a$ is now equal to the $B-B$ bond length $%
a_{B-B}.\;$(Note that for graphene we have the relation $a=\sqrt{3}a_{C-C}$%
). We choose ${\bf a}_{3}\;$along the positive $y-$axis which yields for the
coordinates
\begin{equation}
{\bf a}_{3}=(0\;,\;a)  \label{a3}
\end{equation}
All near neighbour points can then be reached by the vectors $\pm {\bf a}%
_{i}\;(i=1..3).\;$We can then further evaluate the expression for the
structure factor to obtain
\begin{eqnarray}
S({\bf k)} &=&\stackrel{c}{%
%TCIMACRO{\underset{j=1}{\sum }}%
%BeginExpansion
\mathrel{\mathop{\sum }\limits_{j=1}}%
%EndExpansion
}e^{i{\bf k}.({\bf a}_{j})}  \nonumber \\
&=&2\stackrel{3}{%
%TCIMACRO{\underset{i=1}{\sum }}%
%BeginExpansion
\mathrel{\mathop{\sum }\limits_{i=1}}%
%EndExpansion
}\cos ({\bf k}.{\bf a}_{i})  \nonumber \\
&=&2\cos (ak_{y})+4\cos \left( \frac{\sqrt{3}}{2}ak_{x}\right) \cos \left(
\frac{1}{2}ak_{y}\right)  \label{sana}
\end{eqnarray}
As mentioned above, this structure factor is common to both the $QN$ model
and the tight binding result.

\subsubsection{Constant energy surfaces and DOS of the 2D\ boron sheet}

The constant energy contours from the QN model are given in Fig. $3.2$.
Similar to the carbon case, we note that also here so-called 'trigonal
warping effect' \cite{trig} occurs, whereby the constant energy contours
change from the circular behaviour around the center and the $K-$points to a
triangular shape near the $M$ points.

The $DOS_{2D}$ of the $2D$ energy bands can then be calculated from the
expression
\begin{equation}
DOS_{2D}(E)=\frac{2}{(2\pi )^{2}}\int \frac{ds}{\left| \nabla _{{\bf k}}E(%
{\bf k)}\right| }  \label{DOS2D}
\end{equation}
where the line integral is along a surface of constant energy $E$. Returning
to the constant energy surfaces in Fig. $3.2$, we have performed the line
integration in Eq.(\ref{DOS2D}) with $E(k)$ given for the 'model boron
lattice' by Eq.(\ref{QN}) with $c=6$, and the results are depicted in Fig. $%
3.3$. It is worthwhile to compare and contrast the form of $N(E)$ shown
there with the $\pi -$bands for graphite first obtained by Coulson
   \cite{Coulson}
using the same $QN$ model. First, the $\pi -$bands touched in
graphite, at the point where $N(E)=0$, i.e. one had semimetallic behaviour
(both zero gap and zero $N(E_{F}))\;$at the Fermi (or {\it HOMO}) level $%
E_{F}.\;$In contrast, the 'model boron lattice', by counting the $\pi -$%
electrons, is metallic (zero gap and non-zero $N(E_{F})\;$at the {\it HOMO}
level). A van Hove singularity ($vHs)$ arises near the edge of the Brillouin
zone where the energy bands level off due to periodicity. Results from the
tight binding method yielded similar results for all features.

\subsection{1D energy bands of (n,n) i-zigzag boron nanotubes}

Within the chosen orientation for the $x-y$ coordinate system, the chiral
vector ${\bf C}_{h},$ (and correspondingly also ${\bf K}_{1}$) is directed
along the positive $x$-axis for the special case where $m=n.\;$Whereas for
graphene this special choice for the indices leads to nanotubes exhibiting
armchair symmetry, in the case of boron it leads to $i$-zigzag symmetry, as
discussed above. Applying periodic boundary conditions along the
circumferential direction then leads to\ the quantization
\begin{equation}
k_{x,j}=\frac{2\pi }{n\sqrt{3}a}j\;\;\;\;\;\;\;\;\ (j=1,.....,2n)
\label{quantx}
\end{equation}
This then yields for the $i-$zigzag structure factor $S_{j}^{iz}({\bf k)}$%
\begin{eqnarray}
S_{j}^{iz}(k{\bf )} &=&2\cos (ak)+4\cos \left( \frac{\pi j}{n}\right) \cos
\left( \frac{ak}{2}\right)  \label{sk} \\
(j &=&1,...,2n)\;\;\;\;\;\;\ \left( -\frac{\pi }{a}<k<\frac{\pi }{a}\right)
\nonumber
\end{eqnarray}
From this expression, one can immediately write down the corresponding
expression for the $(n,n)$ energybands $E_{q}^{iz}(k{\bf )}${\bf ,} either
in the tight binding approximation or from the Quantum Network model.

Fig. $3.4\;(a)$ shows the energy bands obtained from the $QN$ method for a
boron $(3,3)$ $i$-zigzag nanotube which has from Eq.(\ref{length}) a
diameter $L\approx 5.20a$. Clearly since ${\bf K}_{1}$ is directed along a
symmetry axes of the $2D$ Brillouin zone, we have energy bands symmetrical
around the origin. From tight binding calculations we found a similar set of
bands, with the same general form and degeneracy.

\subsection{1D energy bands of (n,0) e-zigzag boron nanotubes}

We will now go on to derive an expression for the energy bands in the
special case that $m=0.\;$As in the case of graphene this leads to zigzag
nanotubes but as discussed above, we now need to make the extra
specification that we are dealing with equilateral zigzag symmetry .\ In
order to obtain an explicit expression for the energy bands it is helpful to
consider a coordinate transformation to a more suitable reference system.
Following Saito, Dresselhaus and Dresselhaus \cite{nano} we consider a
passive coordinate transformation counterclockwise over $120%
%TCIMACRO{\UNICODE[m]{0xb0}}%
%BeginExpansion
{{}^\circ}%
%EndExpansion
.\;$This implies we now use the $x-y$ coordinate system depicted in their
Fig. $4.3$ $(b)$. This transformation leaves the expression for the
structure factor (\ref{sana}) unchanged as one can easily check, but brings
the chiral vector ${\bf C}_{h}\;$for the $(n,0)$ system under consideration
along the (negative) $y$-axis. Again applying periodic boundary conditions
along the circumferential direction leads to\ the quantization
\begin{equation}
k_{y,j}=\frac{2\pi }{na}j\;\;\;\;\;\;\;\;\ (j=1,.....,2n)
\end{equation}
For the $e$-zigzag structure factor $S_{j}^{ez}(k{\bf )\;}$one then obtains
\begin{eqnarray}
S_{j}^{ez}(k{\bf )} &=&2\cos (\frac{2\pi }{n}j)+4\cos \left( \frac{\pi j}{n}%
\right) \cos \left( \frac{\sqrt{3}}{2}ak\right)  \label{szig} \\
(j &=&1,...,2n)\;\;\;\;\;\;\ \left( -\frac{\pi }{\sqrt{3}a}<k<\frac{\pi }{%
\sqrt{3}a}\right)  \nonumber
\end{eqnarray}
from which one can again immediately obtain the corresponding $(n,0)$\
energybands $E_{j}^{ez}(k{\bf ).\;}$The resulting energy bands from the $QN$
model are plotted in Fig. $3.4\;(b)$. We chose a value for $n$ which yields
approximately a nanotube of the same radius as for the $(3,3)$ zigzag tube,
namely $n=5$ which yields for the circumference $L=5a.\;$

\subsection{1D energy bands for general chiral symmetry}

For a general $(n,m)$ system we go back to the original coordinate system
(as for the armchair nanotubes). The coordinates of the lattice vectors $%
{\bf b}_{1}\;$and$\;{\bf b}_{2}$ in reciprocal space are then given by
\begin{equation}
{\bf b}_{1}=(\frac{2\pi }{\sqrt{3}a}\;,\;\frac{2\pi }{a})\;\ \ \ ,\;\ \;\;\;%
{\bf b}_{2}=(\frac{2\pi }{\sqrt{3}a}\;,\;-\frac{2\pi }{a})  \label{blat}
\end{equation}
which implies the relations
\begin{equation}
{\bf b}_{1}.{\bf b}_{1}={\bf b}_{2}.{\bf b}_{2}=\frac{4}{3}\left( \frac{2\pi
}{a}\right) ^{2}\;\;\text{and}\;\;{\bf b}_{1}.{\bf b}_{2}=-\frac{2}{3}\left(
\frac{2\pi }{a}\right) ^{2}  \label{bin}
\end{equation}
We then obtain for the $x$ and $y$ components of the vector $k\frac{{\bf K}%
_{2}}{\left| {\bf K}_{2}\right| }+\mu {\bf K}_{1}$
\begin{eqnarray}
k_{x} &=&\frac{k}{\left| {\bf K}_{2}\right| }\left[ \frac{m-n}{N}\right]
\left( \frac{2\pi }{\sqrt{3}a}\right) +\mu \left[ \frac{m+n}{N}\right]
\left( \frac{2\pi }{a}\right) \sqrt{3}  \label{kxchi} \\
k_{y} &=&\frac{k}{\left| {\bf K}_{2}\right| }\left[ \frac{m+n}{N}\right]
\left( \frac{2\pi }{a}\right) +\mu \left[ \frac{n-m}{N}\right] \left( \frac{%
2\pi }{\sqrt{3}a}\right)  \label{kychi}
\end{eqnarray}
with
\begin{equation}
\left| {\bf K}_{2}\right| =\sqrt{{\bf K}_{2}.{\bf K}_{2}}=\sqrt{\left[
\left( \frac{m}{N}\right) ^{2}+\left( \frac{n}{N}\right) ^{2}\right] {\bf b}%
_{1}.{\bf b}_{1}-2\left( \frac{nm}{N}\right) {\bf b}_{1}.{\bf b}_{2}}
\label{k2length}
\end{equation}
and the dependence of $N$ on $m$ and $n$ given by Eq.(\ref{D}). For a
general choice of $n$ and $m$, the number of atoms in the unit cell of the
nanotube, and correspondingly the number of $1D$ dispersion relations gets
very large compared to the simpler $(n,n)$ and $(n,0)$ cases and is less
insightfull.

\subsection{Density of States of one-dimensional boron nanotubes}

We now calculate the $DOS$ per boron atom from these $1D$ energy dispersion
relations using the formula
\begin{equation}
DOS(E)=\frac{2}{\left( 2\pi \right) }\frac{\left| {\bf T}\right| }{N}%
\stackrel{N}{%
%TCIMACRO{\underset{\mu =1}{\sum }}%
%BeginExpansion
\mathrel{\mathop{\sum }\limits_{\mu =1}}%
%EndExpansion
}\int \delta \left( E-E_{\mu }(k)\right) dk  \label{dosnano}
\end{equation}
where the integration is over the $1D$ Brillouin zone of the nanotube. The
results are shown in Fig.$3.5(a)\;$and $(b)$ for the $(3,3)$ $i$-zigzag tube
and the $(5,0)$\ $e$-zigzag tube respectively. Obviously since the first
cuts in the $2D$ energy bands of the boron sheet are within the region where
the constant energy contours are circular, irrespective of the direction of $%
{\bf C}_{h}(n,m)\;$we find a similar $DOS$ for the low energy values. On the
whole we find that the $DOS$\ looks similar, the main difference being that
the two peaks which were nearly degenerate for the $(3,3)$ $i$-zigzag tube
are further apart for the $(5,0)$\ $e$-zigzag tube.

All of the boron nanotubes we considered were found to be metallic. Of
course, we must expect also here that a Peierls distortion \cite{nano}
will occur which will then possibly open a small gap in the $DOS$ of the
first energy band but this is beyond the scope of the present study.

\section{'The March model' applied to nanotubes}

Contrary to the density of states, we expect other quantities relating to
the electrons to be only very slightly dependent on the detailed structure
of the carbon or boron frame. To exemplify this point F.E. Leys et al.
\cite{surface} calculated the self consistent field in which the $2s$ and $%
2p$ valence electrons move in an isolated carbon nanotube under the
assumption that the positive charge of the $C^{4+}\;$ions can be smeared out
{\it uniformly} over the surface of an infinitely long cylinder, thereby
neglecting all structure and only retaining the basic cylindrical symmetry
of the problem. The spirit of this assumption goes back to the early work of
March \cite{orima} on tetrahedral and octahedral molecules where, for
instance in $GeH_{4},\;$the external field in which the electrons move was
approximated by smearing out the four $H$ protons over the surface of a
sphere of equal radius to the $Ge-H$ bond length and centered on the $Ge$
nucleus. The inhomogeneous electron gas created by this model external
potential was then treated using the Thomas-Fermi method. In evaluating the
nuclear-nuclear potential energy however the correct geometry of the nuclear
framework was always retained. This method has recently experienced renewed
interest. Thus the work of Clougherty and Zhu \cite{Clou} on $C_{60}\;$%
calculated the equilibrium cage radius for $C_{60}\;$using the method they
termed 'the March-model'. Further studies include those of Despa
   \cite{desp}, also on $C_{60}$, and Amovilli and March \cite{boroncl} on
boron cages. (See also Siringo et al. \cite{Sir}). We will now briefly
summarize the main results of the study of Leys et al. \cite{surface}.

If we take $\Theta (r)\;$to be minus the self-consistent potential $%
V_{sc}(r) $ we obtain for$\;\Theta (r)\;$the differential equation
\begin{equation}
\frac{\partial ^{2}\Theta }{\partial r^{2}}+\frac{1}{r}\frac{\partial \Theta
}{\partial r}=c\Theta ^{\frac{3}{2}}\;\;\;\;\;\;\;\;\ c=\frac{2^{\frac{7}{2}}%
}{3\pi }  \label{tfeq}
\end{equation}
where $r$ denotes the distance from the axis of the tube. This equation was
solved subject to the appropriate boundary conditions and in particular the
surface charge $\sigma $ was taken into account through the boundary
condition at the radius$\;R_{t}\;$of the tube
\begin{equation}
\left( \frac{\partial \Theta _{1}}{\partial r}\right) _{R_{t}}=4\pi \sigma
+\left( \frac{\partial \Theta _{2}}{\partial r}\right) _{R_{t}}
\end{equation}
where$\;\Theta _{1}\;$denotes the solution inside the tube and $\Theta _{2}$
outside the tube. The condition (\ref{gauss}) reflects the discontinuity in
the electric field across the surface charge..

Figs. 4.1 shows our results for the potential $V_{sc}(r)$. The data used in
determining $\sigma $ is mentioned in the captions. \"{O}stling et
   al. \cite{ost} performed full Kohn-Sham calculations on the same model system
and we have used their results for comparison. The electron density $\rho
(r) $ was found to be reasonably well described by $TF$ theory but we
especially found very good agreement for the self-consistent potential $%
\Theta (r),$ both qualitatively and quantitatively. Moreover, in the same
study, \"{O}stling et al. compared their results with those obtained from
tight binding when the exact atomic positions were taken into account and
found only very small differences from their Kohn-Sham results on the model
system. This clearly indicates that we can expect also our results,
following the original procedure of March, to be a good approximation to the
self-consistent field. Obviously the main merit of our work is its relative
simplicity allowing one to obtain analytic results in certain cases, for
instance for the power law behaviour at large $r$ which was found to be
\begin{equation}
\Theta (r)=\left( \frac{16}{c}\right) ^{2}\frac{1}{r^{4}}\left[ 1+\frac{F_{1}%
}{r^{c}}+higher\;order\;terms\right] \;\;with\;\ \ c=2\sqrt{6}-4\;\;
\label{pow}
\end{equation}
and $F_{1}$ an integration constant.

\section{Hetero-nanostructures}

So far we have focussed attention on the homonuclear cases of $B$ and of $C$%
. In this penultimate section, we shall extend these considerations to two
specific heteronuclear systems: (i) $BN$ cages and (ii) fluorinated
buckminsterfullerene. We shall take these in turn below

\subsection{Boron nitride cages: effect of electronegativity difference on
electronic structure}

Let us commence this section\ on $BN$ with two general references to this
area: ($a$) the work of Niedenzu and Dawson \cite{nieden} on
boron-nitrogen compounds and ($b$) the more technologically oriented work on
the synthesis and properties of boron nitride edited by Pouch and
   Alterovitz \cite{Pouch}.

With no electronegativity difference included, Fig. $6$ of ref.\cite{Zhu}
shows the H\"{u}ckel $\pi -$bands of the structures depicted in Fig. $5$ of
the same reference. It is immediately clear from the uppermost and the
lowest parts of their Fig. $6$ that the $\pi -$bands$\;$touch, whereas for
the other $3$ structures depicted in their Fig. $5$ there are energy gaps.
But when electronegativity is included in the H\"{u}ckel treatment (i.e. $%
\alpha _{B}\neq \alpha _{N}),\;$the zero-gap cases with $\alpha _{B}=\alpha
_{N}\;$also have energy gaps: the most important consequence of the
electronegativity difference. Zhu et al. \cite{Zhu} also have used the
so-called extended H\"{u}ckel model, but although the bands change somewhat
in detailed shape the salient feature of the electronegativity as
introducing energy gaps where originally $\pi -$bands touched is confirmed.

\subsection{Fluorinated buckminsterfullerene}

A highly fluorinated fullerene, a $D_{3}$ isomer of $C_{60}F_{48}$, has
recently been investigated by $X$-ray fluorescence spectroscopy
   \cite{fluor}.
The results of such study have been interpreted by quantum
mechanical calculations. In particular, it has been confirmed that the
contribution to the electron density of frontier orbitals, namely the
occupied orbitals with the highest energy, comes from six localized $CC$
double bonds. Evidently, the higher electronegativity of fluorine causes, in
this case, a localization of the $\pi $ cloud of fullerene in the $C-F$ bond
regions,\ which are in fact essentially orthogonal to the carbon cage
surface. The twelve carbon atoms which are not bonded to fluorine thus
determine six isolated $CC$ double bonds. Within a very simple
H\"{u}ckel-like formalism the situation can be simulated by substituting $48$
of the $60$ $p$ functions of fullerene by the same number of $\sigma $ $C-F$
bond functions and also, by reducing the connectivity, namely the bond
order, between all neighbour pair of functions in which one of them is a $%
C-F $ bond orbital. Furthermore, the energy of $C-F$ bond functions must be
shifted to lower energies with respect to the $p$ carbon atomic orbitals.
The resulting molecular orbitals are then split essentially in to three
bands. The lowest energy band is built from $48$ occupied orbitals, which
represent the $48$ $C-F\;\sigma $ electron cloud, the intermediate band
being constructed from $6$ occupied orbitals, representing the $6$ isolated $%
\pi $ $C-C$ double bonds, and, finally, the highest energy band which
includes the $6$ empty $\pi $ $CC$ antibonding orbitals.

\section{Summary and Future directions}

We have stressed the way pentagonal 'defects', present say in a $C_{60}\;$%
cage, can be connected with the curvature introduced into an originally
planar graphitic strip. Using Euler's theorem, a one-to-one correspondence
between the geometry of $B$ cages and $C$ cage structures has been proposed.
This has led us to embed topology and connectivity into a quantum current
network approach to a 'model boron' two-dimensional lattice. The electronic
band dispersion relation $E({\bf k})$ for this lattice has been employed to
obtain (i) constant energy $E$ surfaces in the ${\bf k}=(k_{x},k_{y})\;$%
plane and (ii) the density of states of the $\pi -$electrons. The results
are compared and contrasted with the earlier results of Coulson on a
graphite layer. This 'model boron' layer was then wrapped into a 'model
boron nanotube'. After a discussion concerning chirality and symmetry of
these tubes, the electronic properties were discussed within the zone
folding scheme. As to future directions, generalizations which may prove
feasible of earlier work by Dancz et al. \cite{Dancz}. on a
topologically disordered network are proposed.

\begin{acknowledgement}
N.H.M. wishes to thank Prof M.P. Tosi and the SNS for much hospitality
during a visit to Pisa in 2002. N.H.M. \ is grateful to the Francqui
foundation and in particular to Prof. Dr. Eyckmans for support and
motivation. C.A. wishes to acknowledge financial support from Fondi di
Atenes 2002 (University of Pisa). F.E.L. wishes to thank especially Prof. V.
Popov and Dr. D. Lamoen for helpful and stimulating discussions.
\end{acknowledgement}

\appendix

\section{The Quantum Network model.}

\subsection{Introduction and general formalism}

The idea behind the quantum network ($QN$) model is extremely simple. One
joins each atom to its nearest neighbours, and then treats electrons (though
quantum mechanically, of course) as though they flowed through $1D$ wires as
in an electrical circuit obeying Kirchoff's Laws at every node. This was
first introduced by Pauling \cite{pol} and important later contributions
include those of Ruedenberg and Scherr \cite{rue} who applied the method
systematically to a large group of molecules, Coulson \cite{Coulson} who
was the first to apply the method to periodic systems (namely graphene) and
Montroll \cite{Montroll} who studied a class of model potentials along
the bonds which allow for analytical solutions for the wave functions and
the density of states. In particular, these methods have been applied to
conjugated systems where one assumes that the $\sigma $-electrons form the
framework, and the delocalized $\pi -$electrons move along this network.

If$\;\phi \lbrack j]$ denotes the value of the $1D$ wave function at the $j$%
th node (assuming one has made a suitable numbering of all nodes or atoms)
the system of equations one has to solve is given by \cite{Montroll}
\begin{equation}
F(q,\eta )\phi \lbrack j]=\stackrel{c}{%
%TCIMACRO{\underset{i=1}{\sum }}%
%BeginExpansion
\mathrel{\mathop{\sum }\limits_{i=1}}%
%EndExpansion
}\phi \lbrack j_{i}]  \label{main}
\end{equation}
where $F(q,\eta )$ is a general form factor which contains information about
the (possibly parametrized via$\;\eta )$ potential along the bonds and $%
j_{i} $ ($i=1..c$) denote the index of the $c$ near neigbour nodes (or atoms)

For free electrons along the network (i.e. no potential along the bonds),
one has that
\begin{equation}
F(q,\eta )=c\cos (qa)  \label{ffree}
\end{equation}
where $a$ is the bond length (assumed there is only one) and the energy $E$
is given of course by
\begin{equation}
E=\frac{\hbar ^{2}q^{2}}{2m}  \label{freeenergyrel}
\end{equation}
For certain specific choices of the potential acting along the bonds,
analytic formulae for the structure factor $F(q,\eta ),$ the wave functions
along the wires and the resulting DOS can be derived as was shown for
instance in the work by Montroll \cite{Montroll}.

For large molecules it is convenient to use the matrix formulation of the
problem as was proposed by Ruedenberg and Scherr \cite{rue} (to which we
refer the interested reader for more details). If we define for a molecule
containing $N$ atoms or units the vector
\begin{equation}
{\bf \phi =}\left[
\begin{array}{c}
\phi (1) \\
\phi (2) \\
... \\
... \\
\phi (N)
\end{array}
\right]  \label{eigenvector}
\end{equation}
we can write Eq.(\ref{main}) in the form
\begin{equation}
{\bf F\phi =0}  \label{matnot}
\end{equation}
where ${\bf F}$ is the so-called 'connectivity matrix'. It is an $N\times
N\; $matrix, the structure of which\ can\ most\ easily\ be\ explained\ by\
writing\ it\ down\ for\ a\ simple\ molecule,\ for\ instance benzene.\ If\
we\ number all $C$ atoms in a chosen direction from $1$ to $6$, we obtain
for ${\bf F}$ the matrix
\begin{equation}
{\bf F=}\left(
\begin{array}{cccccc}
-F(q,\eta ) & 1 &  &  &  & 1 \\
1 & -F(q,\eta ) & 1 &  &  &  \\
& 1 & -F(q,\eta ) & 1 &  &  \\
&  & 1 & -F(q,\eta ) & 1 &  \\
&  &  & 1 & -F(q,\eta ) & 1 \\
1 &  &  &  & 1 & -F(q,\eta )
\end{array}
\right)  \label{fben}
\end{equation}
where the omitted elements are all zero. The eigenvalues of this matrix, say
$F_{n},\;$then follow from the secular equation
\begin{equation}
\left| {\bf F}\right| =0  \label{seeq}
\end{equation}
combined with Eqs.(\ref{ffree})\ and (\ref{freeenergyrel}) in the case of
free electrons or from a generalization of Eq.(\ref{ffree}) in the case of a
potential along the bonds.

\subsection{Periodic systems}

For periodic systems equations (\ref{main}) or (\ref{matnot}) can be
considerably simplified. If the vectors ${\bf a}_{i}\;(i=1..c)\;$connect
every lattice point to its $c$ nearest neighbours, Bloch's theorem implies
that
\begin{equation}
\phi \lbrack j_{i}]=\phi \lbrack j]e^{i{\bf k.a}_{i}}.  \label{bloch}
\end{equation}
Then, assuming there is only one atom in the Brillouin zone Eq.(\ref{main})
immediately reduces to
\begin{equation}
F(q,\eta )=S({\bf k})  \label{qnfin}
\end{equation}
with the structure factor $S({\bf k})$ given by
\[
S({\bf k})=\stackrel{c}{%
%TCIMACRO{\underset{i=1}{\sum }}%
%BeginExpansion
\mathrel{\mathop{\sum }\limits_{i=1}}%
%EndExpansion
}e^{i{\bf k.a}_{i}}
\]
and ${\bf k}$ a wave vector of the first Brillouin zone of the periodic
structure. For free electrons along the network one immediately obtains for
the energy bands $E^{QN}({\bf k)\;}$%
\begin{equation}
E^{QN}({\bf k)}=\frac{q^{2}}{2}=\frac{1}{2}\left\{ \frac{1}{a}\cos
^{-1}\left( \frac{1}{c}S({\bf k)}\right) \right\} ^{2}  \label{EQN}
\end{equation}
where atomic units have been used. For a more detailed discussion on the use
of the Quantum Network model to periodic systems we refer the interested
reader to the study of Hoerni \cite{Ho}.

\subsection{More recent related theories: Use of Feynman Propagators}

Let us suppose that at time $t=0$ (i.e. canonical density matrix analogue, $%
\beta =it\;$where $\beta =(k_{B}T)^{-1},\;$with $k_{B}\;$denoting
Boltzmann's constant and $T$ the absolute temperature) the electron can be
located at a particular point, say $O$, on one of the line segments of the
tree. The electron wave function on this line segment for $t>0$ is the
free-particle propagator
\begin{equation}
\frac{1}{\sqrt{4\pi it}}e^{i\frac{y^{2}}{4t}}  \label{gauss}
\end{equation}
where the distance $y$ is to be measured from the initial point $O$. The
electron then diffuses outward from $O$. As usual in the network model, the
wave function is required to vary continuously, but the current divides
equally down the remaining $c-1$ branches, where $c$ denotes the number of
nearest neighbours.\ The wave function $\psi \;$and $\frac{1}{c-1}\frac{%
\partial \psi }{\partial x}\;$evaluated at the node serve as initial
conditions for the wave function itself in the next $c-1$ segments. These
boundary conditions go back to Griffiths \cite{Griff}. But, as Ringwood
stresses, in this situation of an infinite tree the network is simply
connected. The same procedure is to be adopted at the next node and so on.

In fact the simplest procedure analytically is to employ the Green function,
to be calculated, between two points, say $O$ and $P$, distanced
respectively $x$ and $x\prime $ from a node. In the initial line segment the
Green function takes the form \cite{Ring}
\begin{equation}
-\frac{1}{2}\frac{1}{\sqrt{-E}}e^{-\sqrt{-E}\left| y\right| }  \label{gauss2}
\end{equation}
where the energy is taken to be negative, positive energies being obtained
by analytic continuation. Putting $\omega =\sqrt{-E},$ the Green function
between the two points $O$ and $P$ then follows as ($T$ denotes Tree)
\begin{equation}
G_{T}(P,O;E)=\left( e^{-\omega x^{\prime }},e^{\omega x^{\prime }}\right)
Z(b_{N-1})Z(b_{N-2})....Z(b_{1})Z(x)%
%TCIMACRO{\binom{\frac{\omega }{2}}{0}}%
%BeginExpansion
{\frac{\omega }{2} \choose 0}%
%EndExpansion
^{-1}  \label{quèquè}
\end{equation}
where $b_{i}\;$is the length of a segment and $Z(b)$ denotes the matrix
\begin{equation}
\frac{1}{2}(q-1)^{-1}\left(
\begin{tabular}{ll}
$qe^{-\omega b}$ & $(q-2)e^{\omega b}$ \\
$(q-2)e^{-\omega b}$ & $qe^{\omega b}$%
\end{tabular}
\right)  \label{matrix}
\end{equation}

The Green function $G_{L}\;$on the lattice $(L$)\ is then obtained by a sum
over suitable restricted walks $(\gamma )$ \cite{Ring} as
\begin{equation}
G_{L}(x,x^{\prime };E)=%
%TCIMACRO{\underset{\gamma }{\sum }}%
%BeginExpansion
\mathrel{\mathop{\sum }\limits_{\gamma }}%
%EndExpansion
e^{i\alpha (\gamma )}G_{T}(\gamma x,x^{\prime };E)  \label{greenlattice}
\end{equation}
where $\alpha (\gamma )\;$enters the phase of the wave function through
\begin{equation}
\psi (\gamma x)=e^{i\alpha (\gamma )}\psi (x)  \label{wave}
\end{equation}
Ringwood \cite{Ring} uses the above results to recover the (more
directly calculated) density of states in a graphitic layer given by
   Coulson \cite{Coulson}.

\section{Tight binding result for the model boron lattice and relation to
the QN model.}

For the model boron lattice considered above, we have only 1 atom per unit
cell which is assumed to contribute only 1 $\pi -$electron. There is thus
only one band in the first Brillouin zone. The Bloch orbital corresponding
to the translational eigenstate ${\bf k}$\ generated by this atom is given
by
\begin{equation}
\phi _{\pi }=\frac{1}{\sqrt{N}}\stackrel{N}{%
%TCIMACRO{\underset{i=1}{\sum }}%
%BeginExpansion
\mathrel{\mathop{\sum }\limits_{i=1}}%
%EndExpansion
}e^{i{\bf k}.{\bf R}_{j}}\varphi _{\pi }({\bf r}-{\bf R}_{i})
\end{equation}
where $\varphi _{\pi }$ denotes the atomic $\pi -$orbital centered on
lattice position ${\bf R}_{i}\;$and $N$ the number of atoms in the crystal.
The energy $E({\bf k)}$\ is then given by the secular equation
\begin{equation}
\left| H-ES\right| =0  \label{sec}
\end{equation}
where the matrices $H$ and $S$\ are now $1-$dimensional. If we denote the
crystal hamiltonian (which is assumed to depend only on ${\bf r}$) by $\hat{H%
}$ the only matrix element of $H,$ say $\left\langle H\right\rangle $, is
given by
\begin{equation}
\left\langle H\right\rangle =\frac{1}{N}\stackrel{N}{%
%TCIMACRO{\underset{i,j=1}{\sum }}%
%BeginExpansion
\mathrel{\mathop{\sum }\limits_{i,j=1}}%
%EndExpansion
}e^{ik.({\bf R}_{i}{\bf -R}_{j})}\left\langle \varphi _{\pi }({\bf r}-{\bf R}%
_{j})\left| \hat{H}\right| \varphi _{\pi }({\bf r}-{\bf R}_{i})\right\rangle
\label{h}
\end{equation}
If we now only take into account nearest neighbour interactions this can be
rewritten as
\begin{eqnarray}
\left\langle H\right\rangle &=&\frac{1}{N}\stackrel{N}{%
%TCIMACRO{\underset{i=j=1}{\sum }}%
%BeginExpansion
\mathrel{\mathop{\sum }\limits_{i=j=1}}%
%EndExpansion
}\;\epsilon _{\pi }\;+\frac{1}{N}\stackrel{N}{%
%TCIMACRO{\underset{i=j=1}{\sum }}%
%BeginExpansion
\mathrel{\mathop{\sum }\limits_{i=j=1}}%
%EndExpansion
}e^{ik.({\bf R}_{i}{\bf -R}_{j})}\left\langle \varphi _{\pi }({\bf r}-{\bf R}%
_{j})\left| \hat{H}\right| \varphi _{\pi }({\bf r}-{\bf R}_{i})\right\rangle
\label{nn} \\
&=&\epsilon _{\pi }+\frac{1}{N}\stackrel{N}{%
%TCIMACRO{\underset{i=1}{\sum }}%
%BeginExpansion
\mathrel{\mathop{\sum }\limits_{i=1}}%
%EndExpansion
}\stackrel{c}{%
%TCIMACRO{\underset{j=1}{\sum }}%
%BeginExpansion
\mathrel{\mathop{\sum }\limits_{j=1}}%
%EndExpansion
}e^{ik.({\bf a}_{j})}\left\langle \varphi _{\pi }({\bf r}-({\bf R}_{i}+{\bf a%
}_{j})\left| \hat{H}\right| \varphi _{\pi }({\bf r}-{\bf R}_{i})\right\rangle
\\
&=&\epsilon _{\pi }+\stackrel{c}{%
%TCIMACRO{\underset{j=1}{\sum }}%
%BeginExpansion
\mathrel{\mathop{\sum }\limits_{j=1}}%
%EndExpansion
}e^{ik.({\bf a}_{j})}\left\langle \varphi _{\pi }({\bf r}-({\bf R}_{i}+{\bf a%
}_{j})\left| \hat{H}\right| \varphi _{\pi }({\bf r}-{\bf R}_{i})\right\rangle
\end{eqnarray}
where the vectors ${\bf a}_{j}$ connect every lattice position ${\bf R}%
_{i}\; $to its $c$ nearest neighbours and
\[
\epsilon _{\pi }=\left\langle \varphi _{\pi }({\bf r})\left| \hat{H}\right|
\varphi _{\pi }({\bf r})\right\rangle
\]
For the model boron lattice considered here we then obtain for the energy
\begin{eqnarray*}
E^{TB} &=&\epsilon _{\pi }+\left\{ \left[ e^{ik.({\bf a}_{1})}+e^{-ik.({\bf a%
}_{1})}\right] t_{1}+\left[ e^{ik.({\bf a}_{2})}+e^{-ik.({\bf a}_{2})}\right]
t_{2}+\left[ e^{ik.({\bf a}_{3})}+e^{-ik.({\bf a}_{3})}\right] t_{3}\right\}
\\
&=&\epsilon _{\pi }+tS({\bf k)}
\end{eqnarray*}
with
\begin{equation}
t_{i}=t=\left\langle \varphi _{\pi }({\bf r-a}_{i})\left| \hat{H}\right|
\varphi _{\pi }({\bf r})\right\rangle \;\;\;\;i=1..3  \label{tdef}
\end{equation}
since the $\pi -$orbitals are symmetric with respect to rotations in the
plane and all near-neighbour atoms are at equal distance, and
\[
S({\bf k)}=\stackrel{c}{%
%TCIMACRO{\underset{j=1}{\sum }}%
%BeginExpansion
\mathrel{\mathop{\sum }\limits_{j=1}}%
%EndExpansion
}e^{i{\bf k}.({\bf a}_{j})}
\]

On the other hand from the $QN$ model we obtained (in a.u.)
\begin{equation}
E^{QN}({\bf k)}=\frac{q^{2}}{2}=\frac{1}{2}\left\{ \cos ^{-1}\left( \frac{1}{%
6}S({\bf k)}\right) \right\} ^{2}  \label{FEQM}
\end{equation}
The function
\begin{equation}
f(x)=\frac{1}{2}\left\{ \cos ^{-1}(x)\right\} ^{2}  \label{notone}
\end{equation}
is plotted over its domain $\left[ -1,1\right[ \;$in Fig. $B1$. To obtain
correspondence with the tight binding calculation this function should be a
linear function with a negative slope. Except very close to the lower
boundary, we indeed find a linear behaviour for $f(x)$ with a slope of about
$-1.2$ which yields for $t$ the value $-\frac{1.2}{c}\;,$where $c$ is the
near-neighbour coordination number, here equal to $6$. The curve has an
intersection at about $1.2$, which then corresponds to $\epsilon _{\pi }.$
Clearly, apart from a topological similarity, both methods also exhibit
analytical similarities.

\section{boron-nitride systems:}

\subsection{Isolated benzene and borazole levels}

The building block of graphene is benzene (without its $H$ atoms: i.e. a
hexagon of $C$ atoms) while for a boron nitride layer it is a hexagon on
which $B$ and $N$ atoms alternate. It is natural enough therefore to appeal
to a simple treatment of bonding in these two molecules an then to extend
the discussion to graphene and a $BN$ layer. Following the survey of one of
us \cite{low} we treat in this Appendix benzene and borazole by a common
approach to their $\pi -$electron level spectrum, following Roothaan and
Mulliken \cite{root}. We note that one of the objectives of their study
was to treat the ultraviolet spectra of benzene and borazole by the semi
empirical molecular orbital method.

Ignoring interactions between neighbouring ring atoms, one has for the
secular equations for the $6$ molecular orbitals derivable from linear
combinations of $2p_{\pi }\;$atomic orbitals of the ring atoms.
\begin{equation}
\left|
\begin{array}{cccccc}
A & 1 &  &  &  & 1 \\
1 & A & 1 &  &  &  \\
& 1 & A & 1 &  &  \\
&  & 1 & A & 1 &  \\
&  &  & 1 & A & 1 \\
1 &  &  &  & 1 & A
\end{array}
\right| =0\;\;\;\;\text{(benzene)}  \label{benmat}
\end{equation}
where all the omitted elements vanish. Note that this equation is idential
to the secular equation obtained from the $QN$ model and so obviously the
structure of the energyspectrum predicted by both methods will be similar.

For borazole we have similarly
\begin{equation}
\left|
\begin{array}{cccccc}
A^{\prime } & 1 &  &  &  & 1 \\
1 & A^{\prime \prime } & 1 &  &  &  \\
& 1 & A^{\prime } & 1 &  &  \\
&  & 1 & A^{\prime \prime } & 1 &  \\
&  &  & 1 & A^{\prime } & 1 \\
1 &  &  &  & 1 & A^{\prime \prime }
\end{array}
\right| =0\;\;\;\;\text{(borazole)}  \label{boramat}
\end{equation}
For benzene, for example,\ $\;A\;$involves not only the desired energy $E$
but also Coulomb and resonance integrals, and the overlap $S$ between
adjacent $C$ atoms (appropriate generalization for borazole involves
evidently both $B$ and $N$ atomic orbitals).

These secular equations can be reduced (by suitable similarity
transformations) to the form
\begin{equation}
\left|
\begin{array}{cccccc}
A^{\prime } & 2 &  &  &  &  \\
2 & A^{\prime \prime } & 1 &  &  &  \\
& 1 & A^{\prime } & 1 &  &  \\
&  & 1 & A^{\prime \prime } & 1 &  \\
&  &  & 1 & A^{\prime } & 1 \\
&  &  &  & 1 & A^{\prime \prime }
\end{array}
\right| =0\;\;\;\;\text{(borazole)}  \label{boramat2}
\end{equation}
where, to pass from the borazole result to benzene, one has to set $%
A^{\prime }=A^{\prime \prime }=A.$

The MO's corresponding to the roots $A=-2$ and $A=-1$ (twice) in benzene, or
the corresponding orbitals in borazole, are each doubly occupied in the
ground state.

Without going into further details, let us simply state the above in a form
such that one can describe the effect of the $\pi -$electron eigenvalues of
bringing molecules together into a larger compound, the molecular levels
then clearly being broadend into bands. To describe this broadening we
follow Coulson and Taylor \cite{coulor} and rewrite the above formulae
in an explicit manner for borazole
\begin{equation}
(\alpha _{B}-E)(\alpha _{N}-E)-g^{2}(\beta -ES)^{2}=0  \label{expl}
\end{equation}
where the Coulomb integrals are denoted by $\alpha _{B}\;$and $\alpha _{N}$
on boron and nitrogen atoms, respectively, while $\beta $ denotes the
so-called resonance integral.

By way of example, to obtain results for benzene, we merely set $\alpha
_{B}= $ $\alpha _{N}=\alpha _{c}\;$and then we find discrete allowed values
of $g^{2}\;$(corresponding to discrete values of $A$ above) as $g^{2}=1\;$%
(twice) and $g=4$. It is to be stressed that the electronegativity
difference between $B$ and $N$ will be reflected in the difference between $%
\alpha _{B}\;$and $\alpha _{N}$. The fact that boron is more electronegative
than nitrogen is reflected in the inequality $\alpha _{B}>\alpha _{c}>$ $%
\alpha _{N}.\;$We have already emphasized in the body of the text that this
is a crucial point in highlighting an essential difference between the
electronic structure of graphene and of a layer of boron nitride.

\subsection{Broadening of $\protect\pi -$electron energy levels into bands.}

For details of the tight-binding calculation of graphite,the interested
reader must consult the study of Coulson and Taylor \cite{coulor}.
However, the point to be reiterated is that the molecular energy levels,
corresponding to discrete values of$\;g^{2}\;$quoted above, will be broadend
into bands. The detailed nature of the structure of graphene merely gives a
spread of values of $g^{2}\;$that is found to embrace the range $0$ to $9$
after considerable calculation. Defining sums and differences of Coulomb
integrals as
\begin{equation}
E_{0}=\frac{1}{2}(\alpha _{B}+\alpha _{N})  \label{sum}
\end{equation}
and
\begin{equation}
\delta =\frac{1}{2}(\alpha _{B}-\alpha _{N})  \label{diff}
\end{equation}
and introducing also $Z=E-E_{0},\;$which merely shift the zero of energy,
and $\gamma =\beta -E_{0}S$, then one obtains
\begin{equation}
Z=\frac{-2\gamma g^{2}S\pm \left[ 4\gamma ^{2}g^{4}S^{2}+4\left( \delta
^{2}+g^{2}\gamma ^{2}\right) \left( 1-g^{2}S^{2}\right) \right] }{2\left(
1-g^{2}S^{2}\right) }  \label{bigz}
\end{equation}
Denoting the result of adopting the plus sign in Eq.(\ref{bigz}) by $%
Z_{+}(g),\;$then by allowing $g^{2}\;$to embrace the range $0$ to $9$
already quoted, it is a straightforward matter to demonstrate that $%
Z_{+}(g)\;$has its lowest value equal to $\delta ,\;$having a range which is
continuous up to a maximum value while, $Z_{-}(g)$ decreases to its minimum
from the value $-\delta .\;$Thus, for the layer of boron nitride, the $\pi -$%
levels are separated into $2$ sub-bands, with an energy gap $2\delta \;$%
which can be associated directly from the above discussion with the
electronegativity difference between $B$ and $N$.\ In contrast, for
graphene, one must put $\alpha _{B}=$ $\alpha _{N}=\alpha _{c}$ and $\delta $
tends to zero. Hence, instead of the (substantial) energy gap ($\sim 4eV)\;$%
in the $BN$ layer, in graphene the $\pi -$bands touch since $\delta
\rightarrow 0.$

\bigskip

\bigskip

$\bf{Figure Captions}:$

\bigskip

\bigskip

Fig. $2.1$: Unfolded section of a hypothetical $C_{N}\;$carbon
cage (thin lines) with the corresponding $B_{n}\;$boron cage on
top of it (thick lines). Boron cage is constructed by imposing the
constraint that every $B-B $ bond crosses exactly one $C-C$\ bond.
The number of boron atoms is then given by $n=\frac{N}{2}+2.$

\bigskip

Fig. $2.2$: Hartree-Fock values of the $\pi -$electron level spectrum of $%
C_{60}$. The degeneracies are in good agreement with predictions
from H\"{u}ckel theory.

\bigskip

Fig. $2.3$: Hartree-Fock values for the valence electron level spectrum of $%
B_{32}.$

\bigskip

Fig. $3.1$: Model boron lattice. Translational vector $\mathbf{T}$
and chiral vector $\mathbf{C}_{h}\;$defining the unit cell of the
nanotube are
shown. Special choices for the chiral vector $\mathbf{C}_{h}(n,n)\;$and $%
\mathbf{C}_{h}(n,0)$ are shown to lead to isosceles zigzag
symmetry and equilateral zigzag symmetry respectively.

\bigskip

Fig. $3.2$: Brillouin zone of the boron lattice. Constant energy
contours (with values in atomic units) for the quantum network
model are shown. Trigonal warping effect causes deviation from
spherical behaviour near the center and $K$ points towards
triangular behaviour near the $M$ points.

\bigskip

Fig. $3.3$: Density of states (including spin degeneracy) for the
$2D$ boron sheet. Outstanding feature is the van Hove singularity
near the edge of the first Brillouin zone.

\bigskip

Fig. $3.4\;(a)$:$\;1D\;$energy dispersion relations from the $QN$
model for a boron $(3,3)$ $i-$zigzag nanotube

\bigskip

Fig. $3.4\;(b)$:$\;1D\;$energy dispersion relations from the $QN$
model for a boron $(5,0)$ $e-$zigzag nanotube.

\bigskip

Fig. $3.5\;(a)$:$\;DOS$ from the $QN$ model for a boron $(3,3)$
$i-$zigzag nanotube.

\bigskip

Fig. $3.5\;(b)$:$\;DOS$ from the $QN$ model for a boron $(5,0)$
$e-$zigzag nanotube.

\bigskip

Fig. $4.1$: Self consistent field for an isolated single wall
nanotube from the March model. In calculating the surface charge,
the $C-C$ bond length was taken to be 2.68 a.u..

\bigskip

Fig. $B.1$: Plot of the function $f(x)=\frac{1}{2}\left\{ \cos
^{-1}(x)\right\} ^{2}$. To obtain analytic correspondence with the
tight binding result we must have a linear relation and this is
found to be true over most of the domain of $f(x)$.

\bigskip

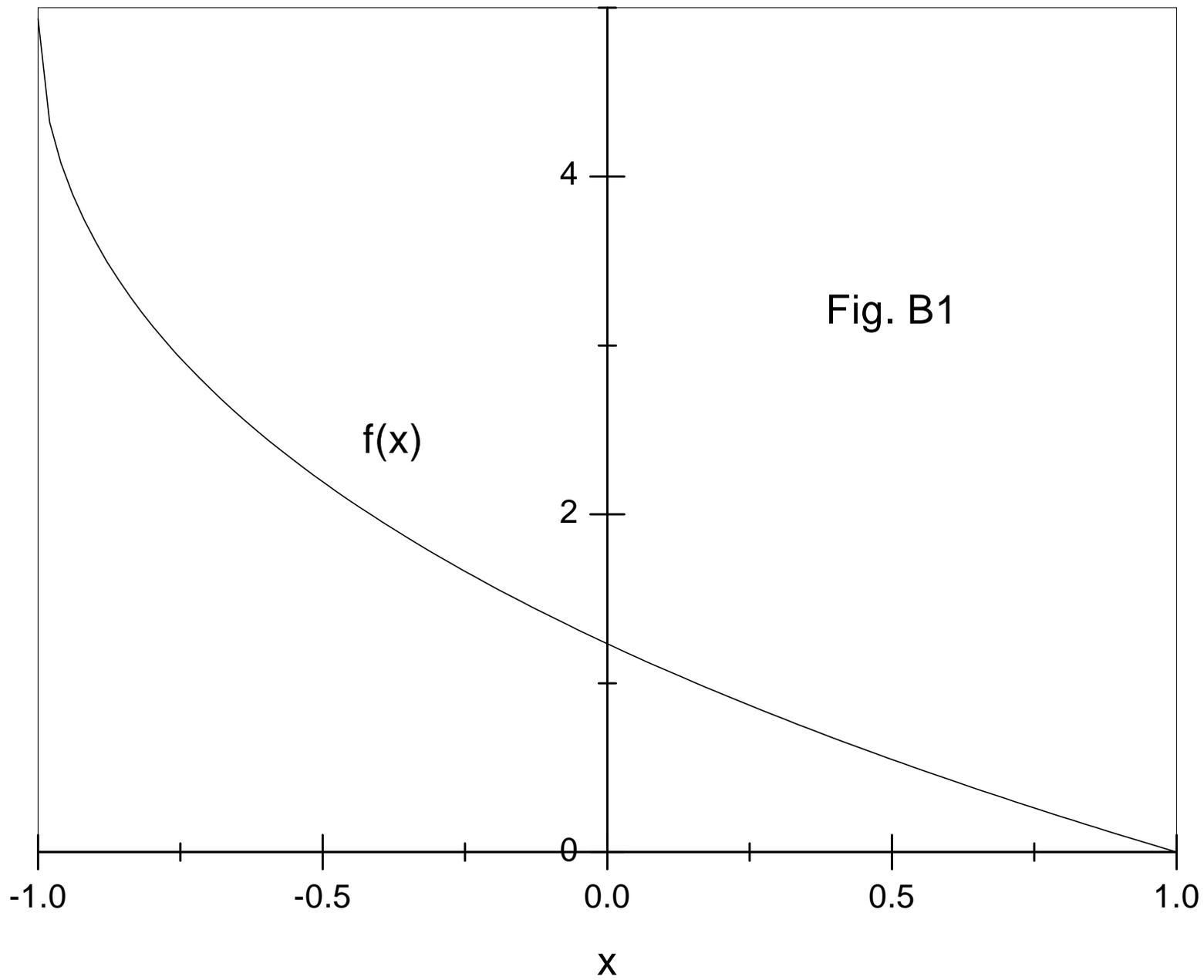
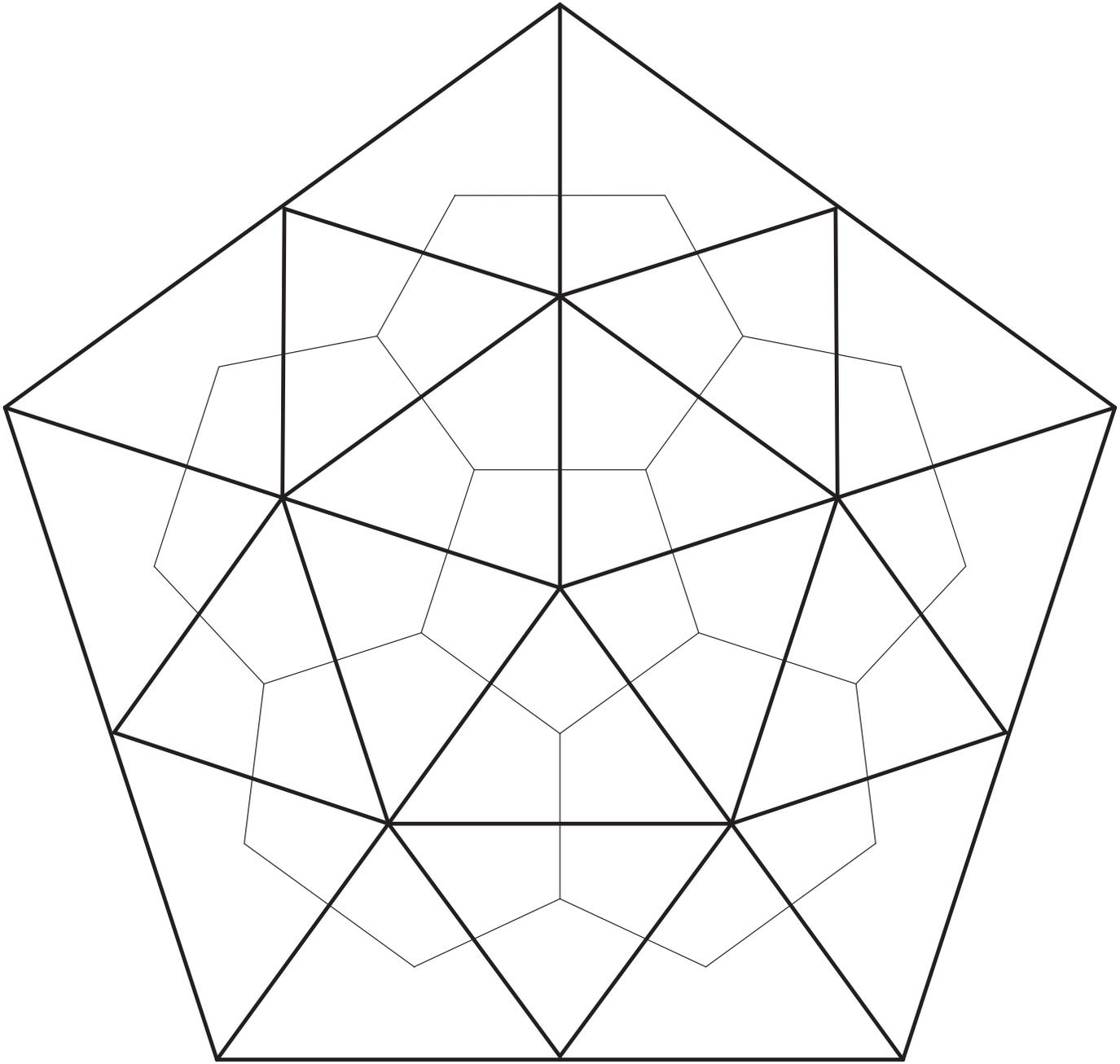
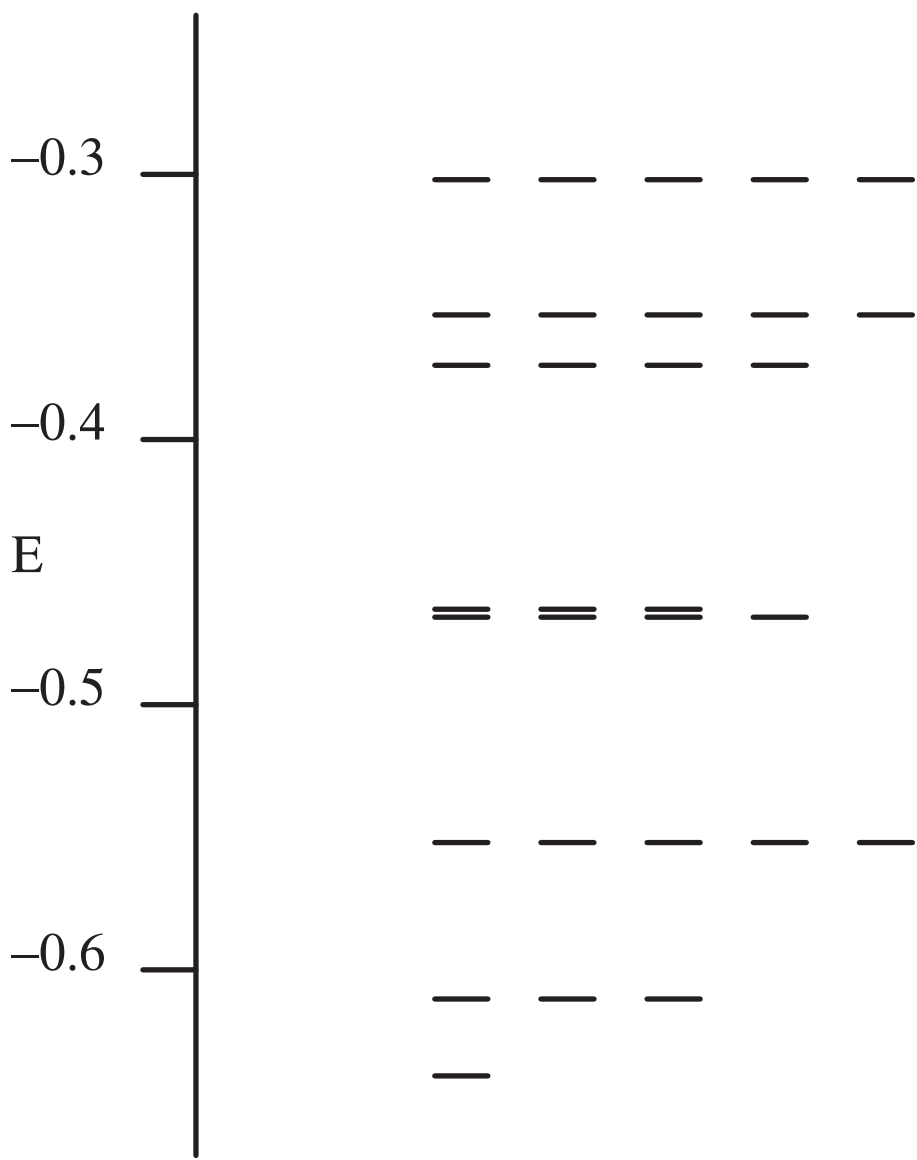
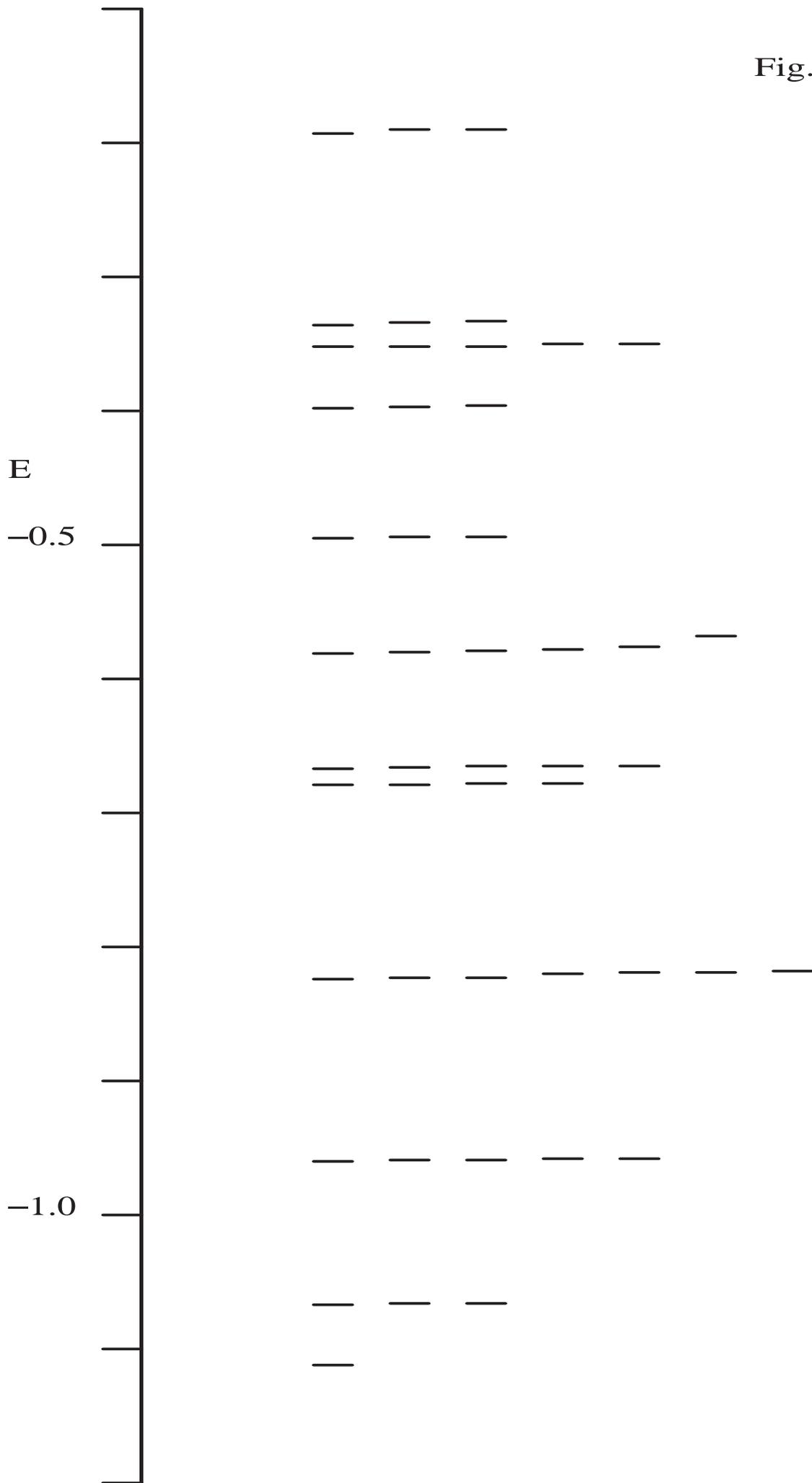
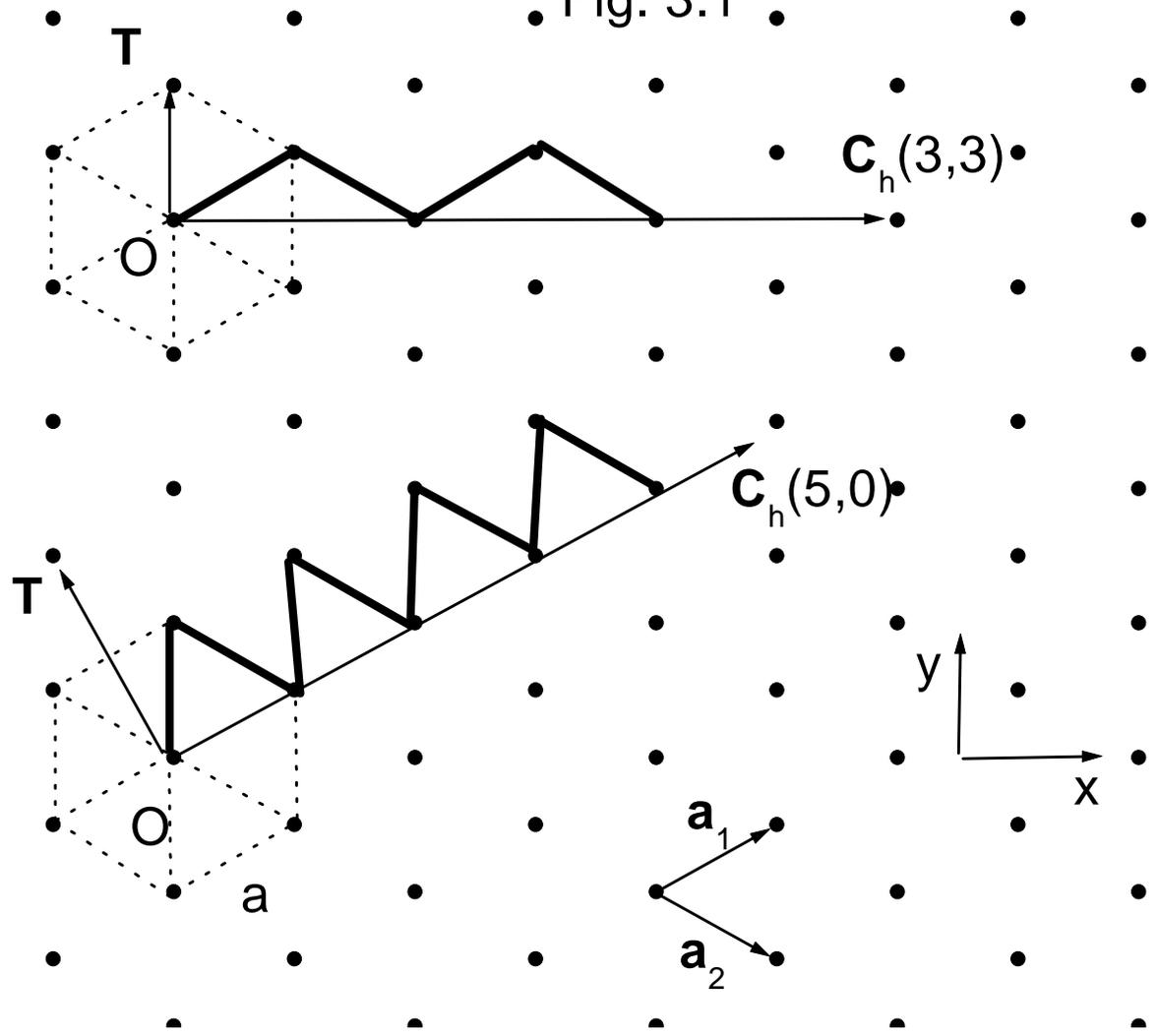
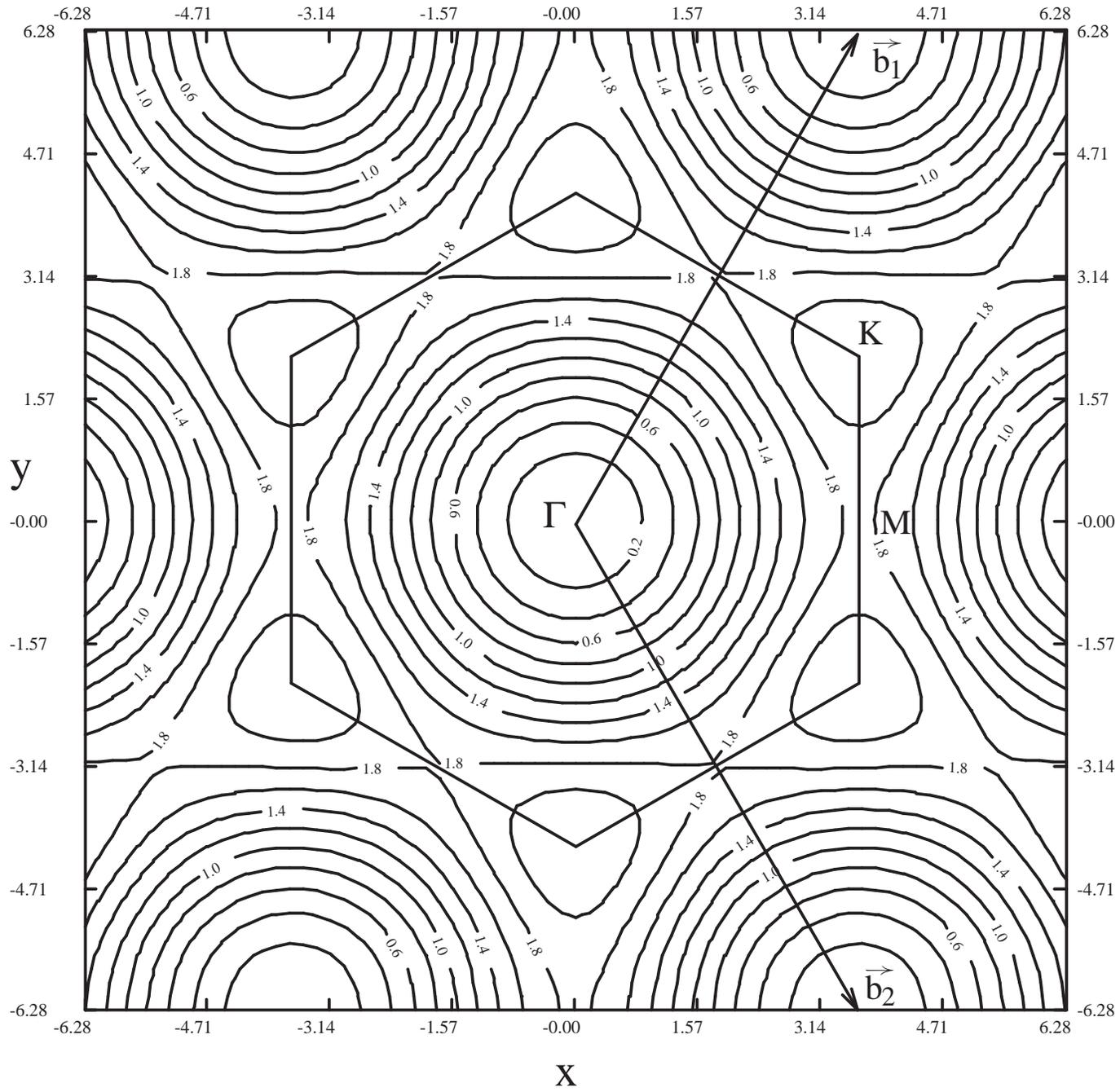
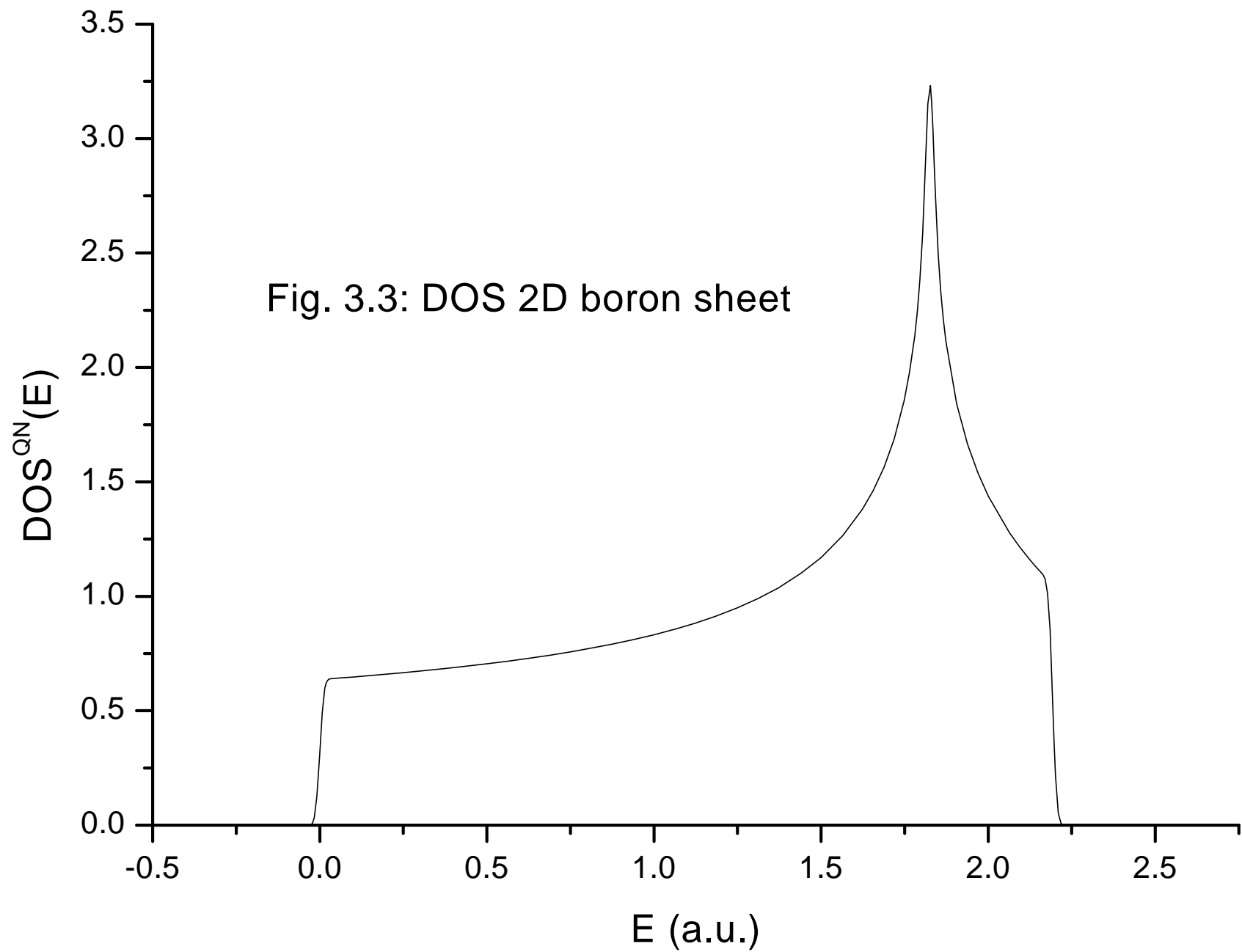
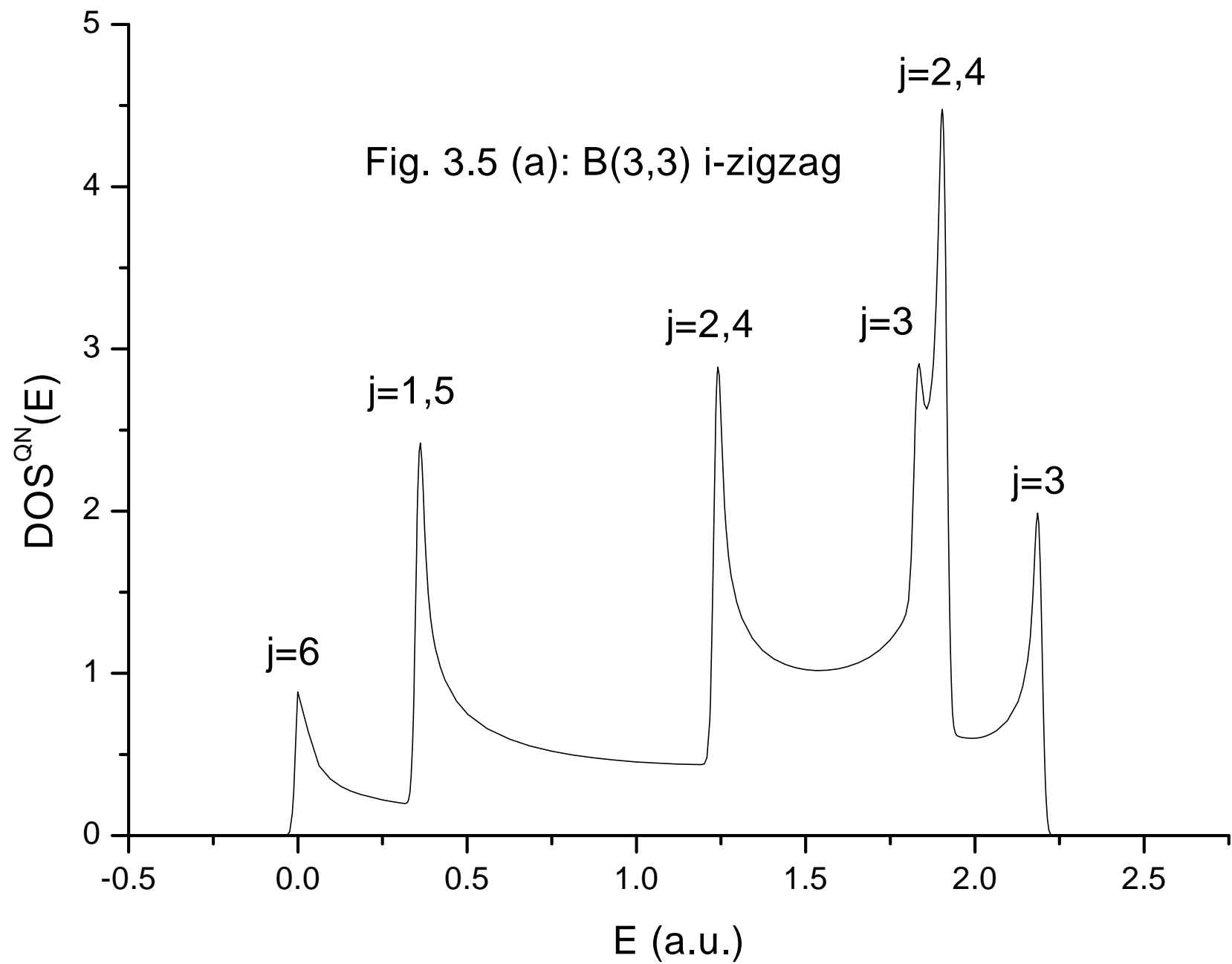
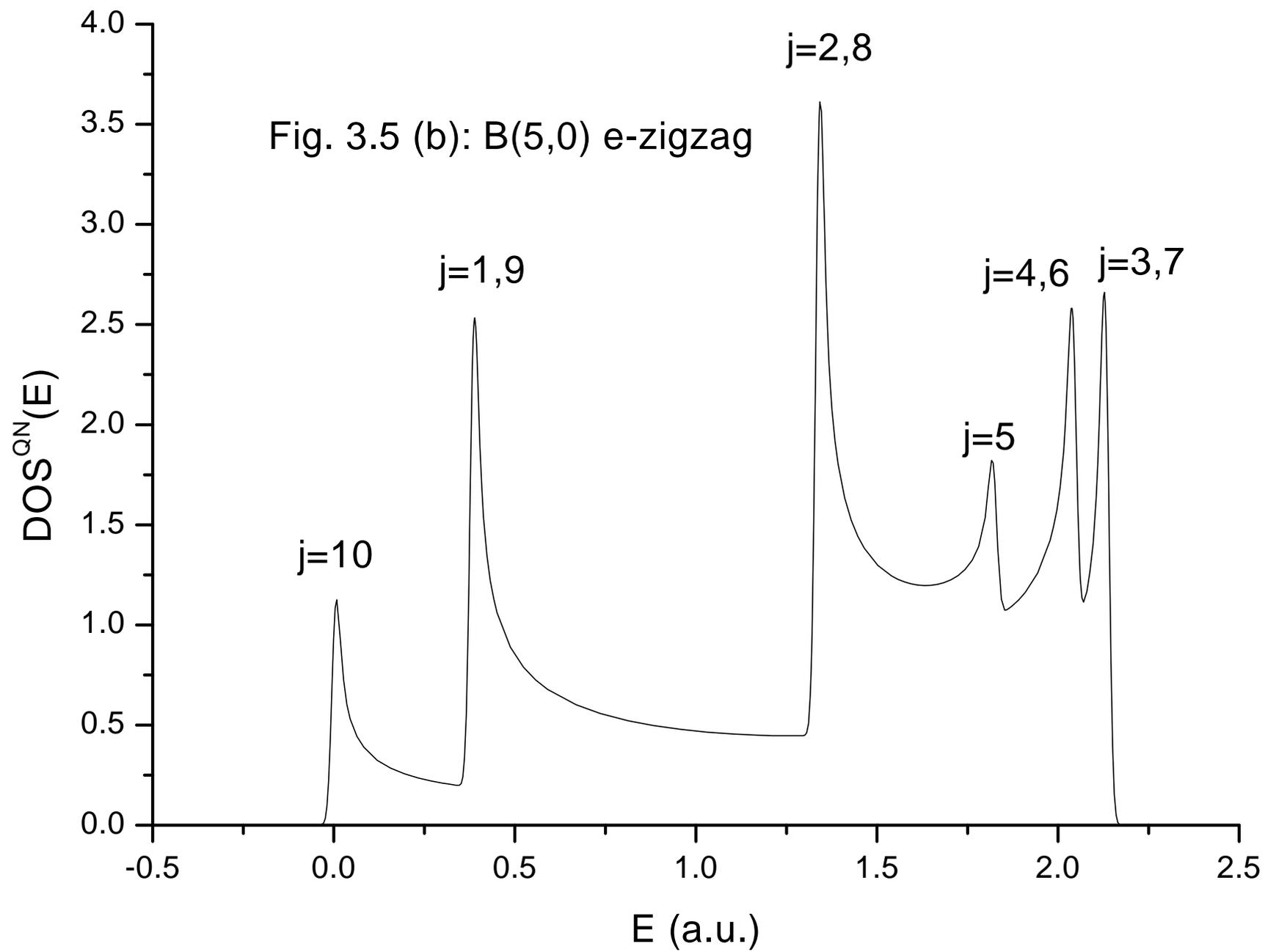
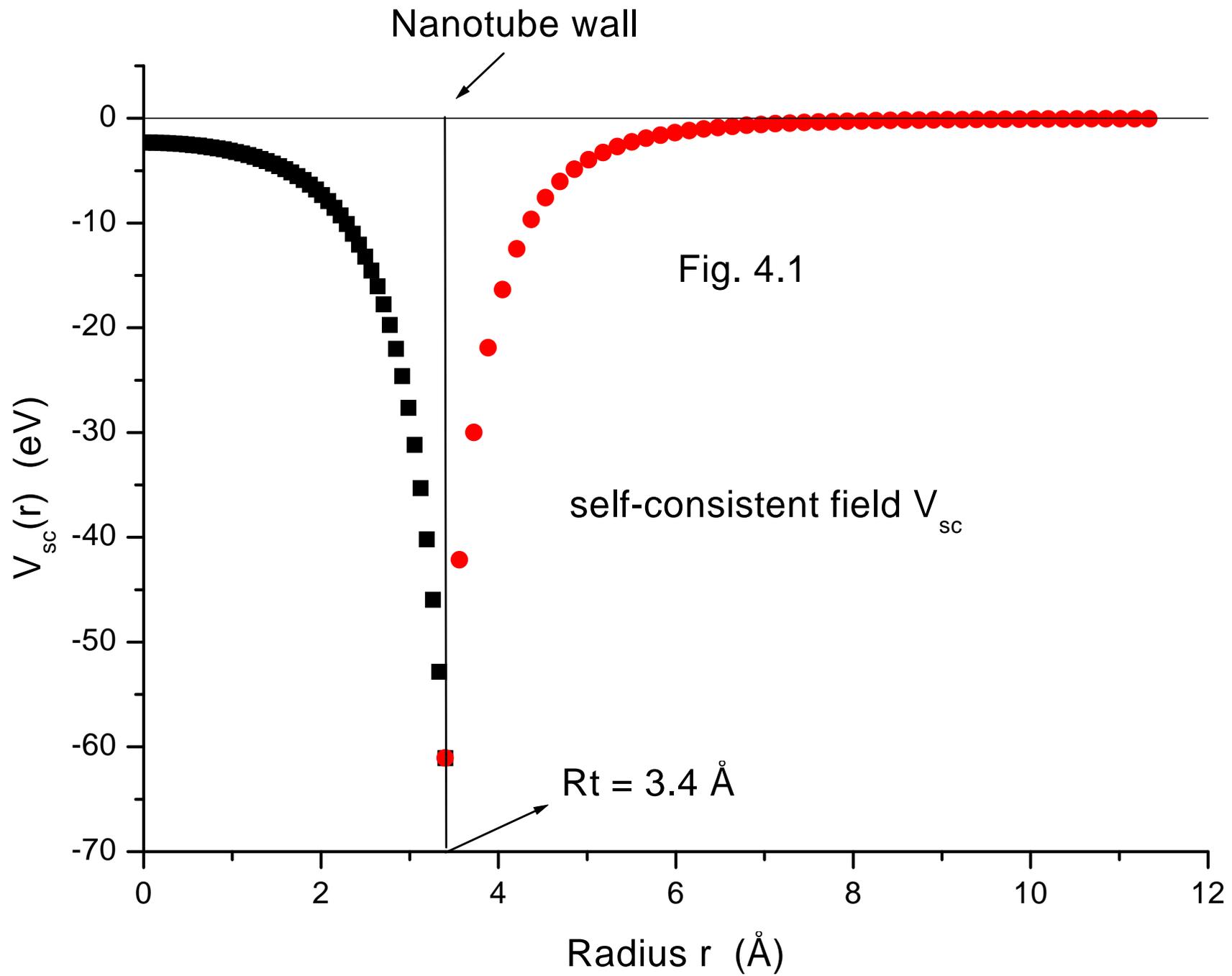
\end{document}